\documentclass[structabstract]{aa}  
\usepackage{graphicx}
\usepackage{txfonts}
\usepackage[]{natbib}
\usepackage{lscape}
\usepackage{longtable}
\newcommand{\ds}{$\rm D_{S}$}    % =obdist
\newcommand{\nm}{$\rm N_{min}$}  % =N_min
\newcommand{\cmd}{B-V \textit{vs} V}

\newcommand{\cc}{B-V \textit{vs} U-B}
\newcommand{\ubq}{U-B \textit{vs} Q}
\newcommand{\qv}{Q \textit{vs} V}

\newcommand{\ebv}{E(B-V)}

\newcommand{\Msun}{$\rm M_{\odot}$}

\newcommand{\halpha}{$\rm H_{\alpha}$}

\begin{document}
   \title{The Young Stellar Population of IC~1613}

   \subtitle{I. A New Catalogue of OB Associations 
\thanks{Based on observations made with 
the Isaac Newton  Telescope operated on the island of La Palma 
by the Isaac Newton Group in the Spanish Observatorio del Roque de los Muchachos 
of the Instituto de Astrof\'{i}sica de Canarias.}
\thanks{Based on observations made with ESO Telescopes at Paranal Observatories
under programme ID 078.D-0767.}
            }

   \author{M. Garcia \inst{1},
           A. Herrero \inst{1,2},
           B. Vicente \inst{1,3},
           N. Castro \inst{1,2},
           L. J. Corral \inst{4},
           A. Rosenberg \inst{1} 
%           P. Stetson \inst{5}
          \and
           M. Monelli \inst{1}
          }

 \offprints{M. Garcia (\email{mgg@iac.es})}

   \institute{Instituto de Astrof\'{i}sica de Canarias, V\'{i}a L\'{a}ctea s/n, E-38200 La Laguna, Tenerife, Spain.
         \and
                Dpto. de Astrof\'{i}sica, Universidad de La Laguna, Avda. Astrof\'{i}sico
                Francisco S\'anchez s/n, E-38071 La Laguna, Tenerife, Spain.
         \and
                Current address: Instituto de Astrof\'{i}sica de Andaluc\'{i}a, CSIC, Apdo.
                 3004, E-18080, Granada, Spain
         \and
                Instituto de Astronom\'{i}a y Meteorolog\'{i}a, Universidad de
                Guadalajara, Av. Vallarta No. 2602, Guadalajara, Jalisco, C.P. 44130, Mexico
%         \and
%                Dominion Astrophysical Observatory, Herzberg Institute of
%                Astrophysics, National Research Council, Victoria, Canada
%%  peter.stetson@nrc-cnrc.gc.ca
             }

   \date{Received February, 2009; accepted 13 April, 2009}

% \abstract{}{}{}{}{} 
% 5 {} token are mandatory
 
  \abstract
   {Determining the parameters of massive stars
is crucial to understand many processes in galaxies
and the Universe, since these objects are
important sources of ionization, chemical
enrichment and momentum.
%Massive stars are amongst the main sources of ionization, chemical
%enrichment and momentum of the Universe.
10m class telescopes enable us to perform detailed 
quantitative spectroscopic analyses of massive stars in other galaxies,
sampling areas of different metallicity.
%a key parameter for their evolution. 
Relating the stars to their
environment is crucial to understand the physical processes
ruling their formation and evolution.
}
   {In preparation for the new instrumentation coming
up on the GTC, our goal is to build
a list of massive star candidates in 
the metal-poor irregular galaxy IC~1613.
The catalogue must have very high astrometric accuracy,
apt for the current generation of multi-object spectrographs.
%so it is suitable for the design of multi-object observations.
A census of OB associations in this galaxy
is also needed, 
to provide important additional information about age and environment of the 
candidate OB stars.
}
   {From INT-WFC observations, we have built an astrometric and photometric
catalogue of stars in IC~1613.
Candidate blue massive stars 
are preselected from their colors. A friends-of-friends algorithm is developed to
find their clustering in the galaxy. 
%\textbf{
While a common physical origin for all the members of the associations
cannot be ensured, 
this is a necessary first step to place candidate OB stars in a population context. %}
}
   {We have produced a deep catalogue
of targets in IC~1613 that covers a large field of view.
To achieve high astrometric accuracy
a new astrometric procedure is developed for the INT-WFC data.
We have also built a  catalogue of OB associations
in IC~1613. 
We have found that they concentrate in the central regions,
%namely the bubbles and the sides of the bar. 
specially in the \ion{H}{ii} bubbles.
The study of extinction confirms that it is patchy,
with local values of color-excess above the foreground value.
}
%   {IC1613 is young, blablabla. All the OBs have the same age
% or not... several episodes of star formation... etc.}

 
        \authorrunning{M. Garcia et al.}
        \titlerunning{OB Associations in IC~1613}
 
   \maketitle
%
%________________________________________________________________

\section{INTRODUCTION}
\label{s:intro}

% ESQUEMA DE LA LINEA DE PENSAMIENTO
%0 Las estrellas masivas son importantes
%1 queremos estudiar las estrellas masivas a baja metalicidad
%para ver como esto afecta a la evolucion y a sus vientos
%2 ic1613 mola porque tiene una metalicidad muy baja
%3 ademas podemos hacer local group cosmology
%4 las estrellas masivas se encuentran en asoc OB y q
%otras ventajas tiene.
%  hasta ahora se habian observado estrellas azules elegidas
%por su magnitud y su posicion en la galaxia.
%al situarlas dentro de una asociacion tenemos mas informacion
%y podemos tb elegir objetos intersantes desde el punto de vista
%evolutivo
%5 asi que en preparacion del gtc nos queremos hacer
%una catalogo de estrellas masivas azules candidatas
%y tb de las asociaciones.
%6 usamos un catalogo que cubre la mayor proporcion
%de la galaxia hasta la fecha y es de los mas profundos
%(el HST no lo cubre todo).
%7 lo dotamos de gran precision astrometrica 
%para que nos sirva para estudios multi-objeto con el GTC.
%8 y nos hacemos el f-o-f para que nos de el catalogo de asoc.

Massive stars are powerful
agents of galactic evolution
because of their stellar winds and UV radiation.
However, these two properties depend strongly on the chemical
composition of the object, making
studies of massive stars in Local Group galaxies of
non-solar metallicity compelling.
Very metal-poor irregulars have an interesting add-on,
local cosmology. 
By studying their massive stars
and OB associations we can learn how
star formation and evolution works at low metallicity, and infer a template
for the deeper Universe. 

To exploit the improved capabilities that will be available
with the \textit{Gran Telescopio Canarias} (GTC),
we are building a list of candidate blue
massive stars in galaxies of the Local Group.
Spectroscopic observations of massive stars
outside the Milky Way and the Magellanic Clouds
require multi-object configurations since
exposure times are long.
Targets are often selected based on their magnitude 
(favouring the brightest candidates)
or because of their location, constrained 
by the multi-object spectrograph used for the observations
(see for instance \citet{Fal07,Cal08}).
However, more interesting candidates
can be found if the target selection criteria 
are not only based on photometry or location
but also on the nurturing population.
Moreover,
OB associations are 
an ideal test bench to study 
the evolution of massive stars
because of their expected small age spread
and homogeneous chemical composition (see e.g. Sim\'on-D\'{\i}az 2009, in prep.).

In this work we study the  magellanic irregular
IC~1613 (Im BV, \citet{FT75}), 
%IC1613 no esta muy claro si esta free-floating in local group 
% (GHB02) or if it belongs to M31 subgroup (McI06).
%  MKJ07: IC1613 is perhaps in a kinematical stream.
because of its proximity (distance modulus DM=24.27 \citet{Dal01}),
low  stellar density and highly resolved population,
very low foreground reddening (\ebv=0.02 \citet{LFM93}, 0.03 \citet{S71} )
and most importantly its low metallicity.
The study of the emission line spectra of some \ion{H}{ii} regions yielded 
0.04 to 0.08 Z$_{\odot}$ \citep{T80,DK82,DD83,PBT88,KB95},
and the stellar abundances of B supergiants range
from $12 + \log O/H=$ 7.80 to 8.00 \citep{Fal07}.

The stellar component of IC~1613 is known to extend little
farther than 16$\arcmin$.5 from its centre \citep{BAG07}.
The optically brightest part is well separated into 2 sides
by a central strip of \ion{H}{i}, usually referred to as 'the bar'.
%\textbf{
The bar is the NE ridge of a hole of the \ion{H}{i} distribution where the SW side
of the galaxy is enclosed \citep{Sal06}.
Star formation is ongoing at the two sides of the
bar and at the SW part of the \ion{H}{i} cavity, %}
but it is more intense on the NE lobe of the galaxy
where spectacular  giant \ion{H}{ii} shells
%similar to those of LMC, SMC, M31 and M33 - 
blown by OB associations \citep{MCW88,VGal01} are seen.
The galaxy has one of the few known WO of the Local Group
(WO3 \citet{KB95}, discovered by \citet{DR82})
and a supernova remnant discovered by \citet{Dal80}.

\citet{S71} presented the first catalogue of 
targets in IC~1613,
based on Baade's work on photographic plates.
Subsequent photometric studies with ever improved detectors 
analysed also the galaxy's population from color-magnitude diagrams
\citep[][ among others]{H78,H80,F88,Hal91,TG02,BAG07,BDA07}.
The DM is well known from numerous studies of Cepheids and RR~Lyrae
(see for instance \citet{Ual01}).
The youngest stars (age $\sim$ 10 Myr)
%concentrate in an torus of radius r$\sim$4\arcmin~ \citep{BAG07},
are patchily arranged
on the uniform underlying distribution of red stars \citep{F88,BDA07}.
In the bubble region, the young population can be split into 2 
age groups of 5-20 Myr and 200-300 Myr  \citep{Gal99}. 
The red supergiant stars in the region
are massive (20-25\Msun) and belong to the younger age group \citep{Bal00}. 

Spectral types have been determined for the brightest stars
of the galaxy 
%\citep{Hu80,Fal07},
and some blue stars in the active bubble region. %\citep{Lal02}. 
The earliest known star is an O3-O4V \citep{Fal07}.
Inside the \ion{H}{ii} super-bubbles, one Of star and three O giants
were found by \citet{Lal02}, but a larger population of O stars
is expected: to reproduce the observed far infrared emission by dust,
it must absorb the equivalent UV emission of about 9 O6-type stars \citep{MCW88}.
Recent quantitative spectroscopic analyses
produced the stellar parameters of %three M-supergiants \citep{Tal07} and 
a sample of B supergiants \citep{Fal07}.
% Humphreys 1980 also provides spectral types, but 
%   does not give coordinates, only the sandage names
The upper main sequence seems to be well populated but
the census of early-type massive stars is not yet complete.

The ultimate goal of our project is to provide a comprehensive
characterization of the massive stellar
population of IC~1613.
In preparation, this paper presents
a deep photometric catalogue (down to V$\sim$25)
that extends well beyond the visually brightest
centre of the galaxy, covering
a total field of view of $\rm 34 \arcmin \times 34 \arcmin$.
The catalogue is provided with a very accurate
astrometric calibration, suitable
for follow-up multi-object spectroscopy with fiber-based instruments.
To understand the population of blue massive stars, 
we have built a new list of OB associations
using our automatic friends-of-friends code \citep[after ][]{B91}.
\textit{The improvement with respect to previous similar works is the input
photometric catalogue, that provides a better coverage of the
galaxy, reaches a fainter V-magnitude than previous ground-based works
and has an improved astrometric solution.
%and the selection criteria for OB stars
%based on the extinction-free Q parameter.
}

The observational dataset and the reduction process
are described in Section~\ref{s:OBS}.
The astrometric solution is provided in Section~\ref{s:astrometry}.
Our code for automatic search of OB associations is described
in Section~\ref{s:code}, and its application to our IC~1613 catalogue 
of candidate OB stars in Section~\ref{s:run}.
The resulting list of OB associations is presented in Section~\ref{s:cat}.
%, and outstanding individual associations are discussed in Section~\ref{ss:comments}.
In Section~\ref{ss:qcfa} we discussed how a different input catalogue
of stars may alter the results.
A comparison with previous catalogues is provided in Section~\ref{ss:comp}.
Finally, we present our conclusions in Section~\ref{s:conclusions}.

\section{OBSERVATIONS AND REDUCTION PROCESS}
\label{s:OBS}

The observations were carried out on the 2.5m
Isaac Newton Telescope (INT) using the 
Wide Field Camera (WFC).
The WFC consists of 4 chips (each one is a thinned EEV CCD,
with $\rm 2048 \times 4096$ pixels) that
together cover a field of $\rm 34 \arcmin \times 34 \arcmin$ 
%(therefore covering the whole galaxy) 
and has a pixel scale of 0.33\arcsec/pixel.
The log of the observations is provided in Table~\ref{T:log}.
They were structured in 3 sets
that included 300s exposures taken with filters BVRI,
and 600s exposures with filter U.
Each set differs from the other on the pointing coordinates:
second and third are offset by $\rm \pm$ 30\arcsec~ in declination
so no star was lost because of the inter-chip gaps or CCD deffects.
An additional short exposure of 60s was taken for all
the filters in the first set, to properly acquire
bright sources.

   \begin{figure*}
   \centering
   \includegraphics[width=16.0cm,angle=0]{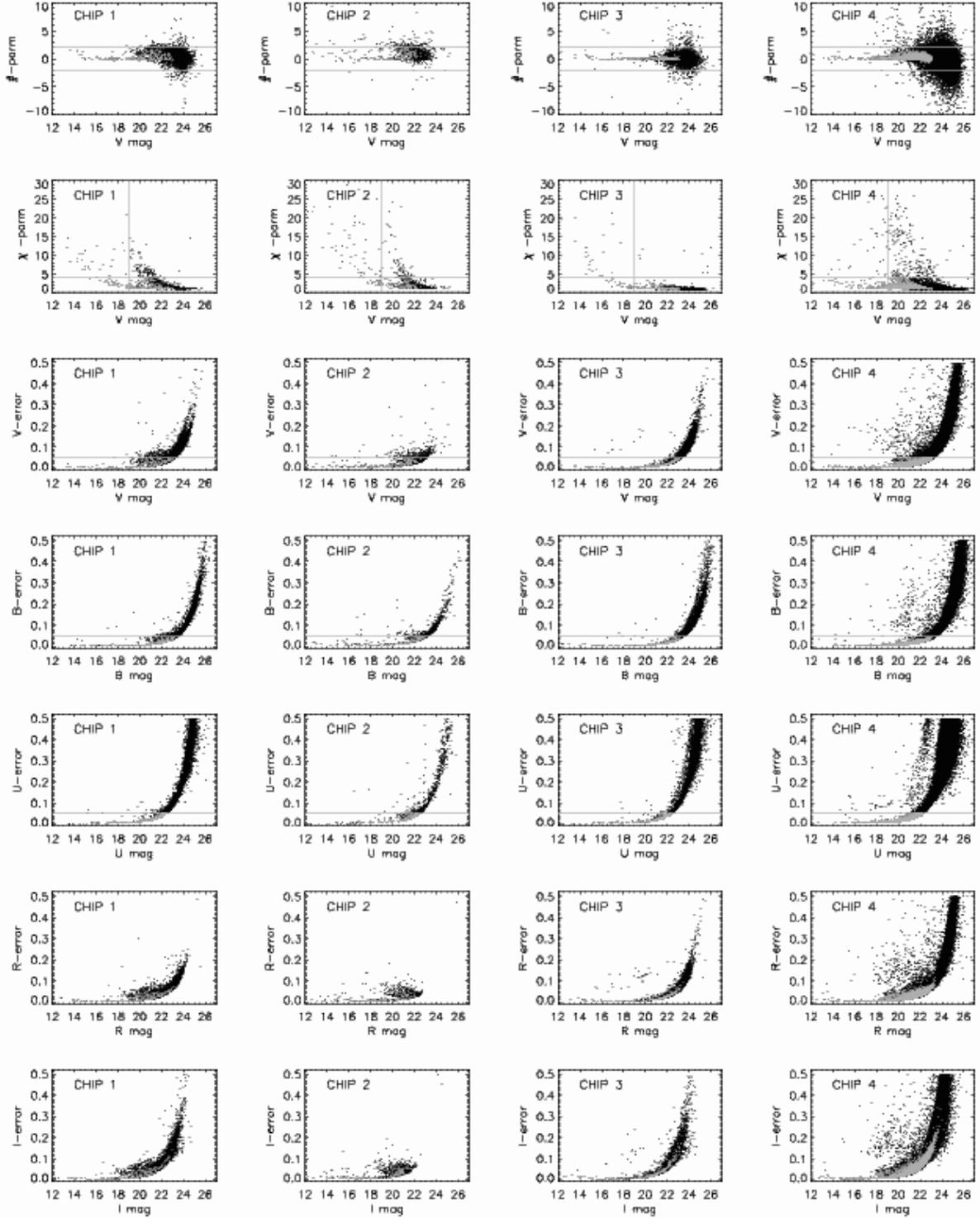}
      \caption{%\textbf{
Photometric accuracy and goodness of DAOPHOT
fit-to-PSF parameters %} 
of targets in our IC~1613 catalogue (black dots)
as a function of magnitude.
Sources registered at different chips
of the WFC are plotted in different columns.
The catalogue's faint limit is V=25. 
V-magnitude errors for $\rm V \leq 23 $ stars are smaller than 0.1.
Grey dots  are the input set of targets for the search of OB associations.
%\textbf{
The horizontal lines in the UBV-errors, \#- and $\rm \chi$-parm. %} 
plots represent
the quality requirements explained in Sect.~\ref{sss:Qchart}.
%For targets brigher than V=19 we admit $\rm \chi$ up to 30. 
%The $\rm \chi$~plot is repeated at a different scale. 
         \label{F:phot}}
   \end{figure*}
%\cleardoublepage
%\newpage

The S/N achieved for an apparent magnitude of V=23
is:
21 / 52 / 37 / 27 / 14 for UBVRI respectively.
%(calculated for 0.6'' seeing,
%airmass=1.3, dark night, 300s for BVRI, 600s for U -and then multiplied
%by sqrt(3)).
As we show in the next Section,
the maximum V-magnitude error for a V=23 star is 0.1.
%\textbf{
The average full-width half-maximum (FWHM) of point sources is
3.5 / 3.2 / 2.8 / 2.6/ 2.6 pixels for UBVRI i.e., 
0.86\arcsec-1.16\arcsec.
At the distance of IC~1613, 1\arcsec~ is equivalent to 3.5 pc.
%}

\begin{table}
\caption{Log of INT-WFC observations of IC~1613}
\label{T:log}
\centering
{\small
\begin{tabular}{c c c c c c c c}
\hline
\hline
 RA         &  DEC     & \multicolumn{5}{c}{T $ \rm _{exp} \left[ \times 100 s \right] $} &  Date \\
 (J2000.0)  & (J2000.0)&  U  &  B  &  V  &  R  &  I  &     \\
\hline
 01:04:48.4 & 02:07:10 & 0.6 & 0.6 & 0.6 & 0.6 & 0.6 & 2006/07/30 \\
 01:04:48.4 & 02:07:10 &  6  &  3  &  3  &  3  &  3  & 2006/07/30 \\
 01:04:50.4 & 02:07:40 &  6  &  3  &  3  &  3  &  3  & 2006/07/30 \\
 01:04:46.0 & 02:06:35 &  6  &  3  &  3  &  3  &  3  & 2006/07/30 \\
\hline
\end{tabular}
}
\end{table}

Complementary VLT-VIMOS images were used for
inspection purposes. The observations
are the pre-imaging part of programme 078.D-0767,
Principal Investigator A. Herrero.

\subsection{Photometric Reduction}   %Matteo ________________________________
\label{ss:matteo}

%\textbf{
The photometric reduction was kindly performed by P. Stetson. %}
The data were first corrected for non-linearity,
following the prescriptions given in the web page\footnote{http://www.ast.cam.ac.uk/$\sim$wfcsur/} of the WFC.
Standard IRAF\footnote{IRAF is 
distributed by the National Optical Astronomy Observatory, which is operated by the Association of Universities for 
Research in Astronomy, Inc., under cooperative agreement with  the National Science Foundation.}
routines were then used for bias
and flat field correction \citep{valdes}. 
No defringing mask was applied to the I band images since no reliable map
could be obtained from this dataset.
 
The source extraction was performed on the four chips independently.
The DAOPHOT/ALLFRAME software package \citep{stetson-dao,stetson-allf}
was used according to the  following protocol.
A first search for stellar sources
and aperture photometry was performed on each individual exposure. %frame (exposure). 
To build the list of stars suitable to model the point-spread function (PSF),
bright and isolated stars 
with small photometric errors and good shape parameter 
( %\textbf{
{\it sharpness}, hereafter \#-parm. %}
)
were selected from the catalogue of extracted sources.
An average of 100 to 200 stars were picked per chip, spread throughout its surface
to properly model the PSF spatial variations.

{\it Allframe} then requires two input
files: a list of stars and
the geometric transformation needed to match different frames. 
The latter was estimated using {\it daomatch}/{\it daomaster}.
A new input list of stars was extracted from the
median-stacked image (built from all exposures and all filters).
{\it Allframe} was run once again.
Adopting the more accurate stellar centroid
and magnitude determinations by {\it allframe} enables (i) a better model
of the PSF, (ii) a refinement of the geometrical transformation, and (iii) a better
median image.

As a by-product, {\it allframe} creates individual residual images after source
extraction. Visual inspection of the median stack of these, revealed
that many stars
were missed in the first search, especially in the central region of the galaxy where
crowding is quite severe.
A second search of stars was performed and the 
new targets were added to the input list of stars of {\it allframe},
which was run a second time with updated PSF and geometrical transformation.
This loop ({\it allframe} -- inspection of the median of the subtracted images
-- update of input list) was iterated a third and a
fourth time for chip~4, that contains the bulk of IC~1613.

Finally, the photometric calibration was performed using local standards from
P. Stetson's catalogue
\footnote{http://www3.cadc-ccda.hia-iha.nrc-cnrc.gc.ca/community/STETSON/standards/}.

The resulting catalogue has almost 85,000 targets.
It will be publicly available on-line, with an accurate 
astrometric calibration for the
target coordinates (see Sect.~\ref{s:astrometry}).
The photometric depth and typical accuracy of the catalogue 
can be seen in Fig.~\ref{F:phot}.

The expected foreground contamination in the central
$\rm 15\arcmin \times 15 \arcmin$ of IC~1613 is $\sim$400
stars with $\rm V \in [16,24]$ and B-V$<$0.8
(\citet{RB85} from the Milky Way model of \citet{BS80}).
This region, covered mainly by Chip~4 and marginally with
Chips~1 and 3, contains approximately 23,500 objects of our catalogue
with the same magnitude and color-range.
Foreground contamination is therefore expected to be negligible
in the centre of the galaxy.
Chip~2 covers the outskirts.
It is the least populated and contains only 235 entries in the listed
magnitude and color range.
%($\rm \sim 23 \arcmin \times 12 \arcmin$)
Even though a large 
population of RGB stars is expected in this location
\citep{BDA07},
Chip~2 is likely dominated by foreground stars.

%Chip~2 is the least populated. 
%Although it is the farthest from the bright optical center of
%the galaxy (21\arcmin), there is still an important population of RGB stars
%which may extend as far as the \ion{H}{i} envelope \citep{BDA07}.
%However, foreground contamination may also be significant.
%According to the estimation of Galactic foreground stars towards
%IC1613 of \citet{RB85}, that uses the \citet{BS80} model for the 
%Milky Way, we may expect
%$\sim$400 foreground stars with V in the interval $\rm [16,24]$, and B-V$<$0.8
%in the central $\rm 15\arcmin \times 15 \arcmin$.
%In this region, our catalogue has approximately 23,000 objects
%with this magnitude and color-range, and we expect foreground
%contamination to be negligible.
%However, only 235 objects are present in Chip~2 
%($\rm \sim 23 \arcmin \times 12 \arcmin$), 
%which we expect to be dominated by foreground stars.

%__________________________________________________________________

\section{ASTROMETRIC CALIBRATION FOR WFC IMAGES}
\label{s:astrometry}

The importance of an accurate astrometric solution for
any kind of source catalogue is often underestimated.
Firstly, it is essential for cross-identification with other existing
and future databases, allowing the combined use of complementary data.
Secondly, it enables follow-up observations with instruments demanding high
position-accuracy, 
such as multi-object spectrographs.
Additionally, instruments with a large field of view
(such as the INT-WFC)
%The INT-WFC has a very large field of view
%($34\arcmin \times 34\arcmin$)
suffer from large geometric distortions.
Our astrometric calibration 
ensures
that the instrumental defects on the target
positions are totally removed.

%As mentioned earlier, the WFC at the INT
%consists of four CCD chips of $\rm 2048 \times 4096$ pixels
%each, with a pixel scale of $0.33\arcsec$ pixel.
%Its wide field of view ($34\arcmin \times 34\arcmin$)
%introduces large geometric distorsions
%that must be removed prior to any astrometric calculations.

\begin{figure}[b!]
\centering
\includegraphics[width=0.4\textwidth]{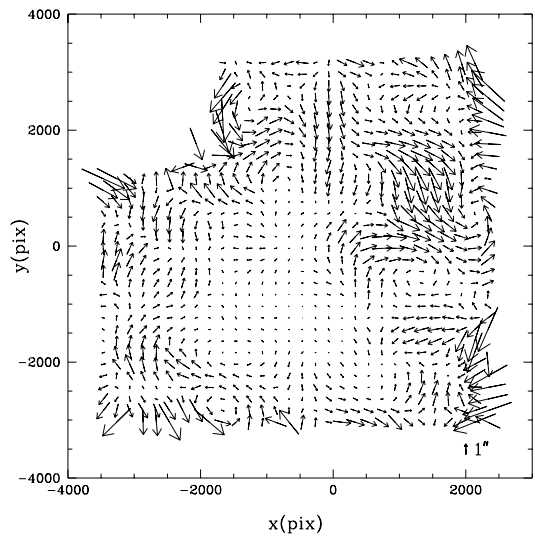}
\caption{Stacked position residuals of the complete INT-WFC mosaic.
The arrows represent the difference between
the positions listed by 2MASS and our catalogue,
for the reference stars.
The scale vector of 1\arcsec size is plotted on the lower right corner.
A radial correction plus a global cubic polynomial solution has
been applied to our catalogue.
The positional differences vary
as a function of the mosaic coordinates x and y.
The fixed pattern shows that instrumental distortions
have not been completely removed.}
\label{fig:radial}
\end{figure}

Typically, the transformation from instrumental
coordinates into celestial coordinates
is accomplished by describing the instrument focal plane with
a polynomial model. An existing reference catalogue is used to
determine the parameters of the model, i.e. the coe\-fficients of
the polynomial.
A significant overlap of stars between the observations
and the reference ca\-ta\-logue
is needed in each exposure, to ensure that the model is well constrained. 
The solution is then applied to all (x,y) detections,
yielding celestial coordinates for all observed objects 
and not only for the reference stars.

The routine we used for astrometric calibration
is fully automatic and does not require a pre\-vious 
manual cross-identification of the observed stars
in the reference catalogue.
The programme first makes a pre\-li\-mi\-na\-r
cross-identification matching geometrical patterns of
both input lists of stars
(the measures and the reference one)
in the frame's coordinate system.
A linear transformation is calculated and 
applied to all stars. 
%Based on the transformation solution, the
%routine also calculates an improved estimate of the tangent point,
%which it updates. 
A new cross-identification is made,
this time  only searching for positional matches
with a tolerance of $0\farcs5$.
This way, a more complete list of common stars 
is achieved.
They are used as re\-fe\-ren\-ce stars to calculate the final
solution, a full cubic polynomial transformation.
A more detailed description of the procedure is provided in
\citet{CL06}.

In this work, the reference positions were taken from the 2MASS Catalogue
\citep{Cal03} whose reference system is already on the
International Celestial Reference System (ICRS).
The 2MASS Catalogue does provide a dense reference list of targets with precise
positions. Proper motion correction was not necessary since our observations and 2MASS'
are contemporary.
Since 2MASS is an infrared survey,
we cross-correlated our catalogue's R magnitude with the J magnitude of 2MASS.
The starting point was the instrumental (x,y) positions derived from the PSF fitting
described in Section~\ref{ss:matteo}.

%\subsection{Reference star identification and polynomial solution}

\subsection{Mosaic System and Distortion Removal}

\begin{figure*}[t!]
\centering
\includegraphics[width=0.85\textwidth]{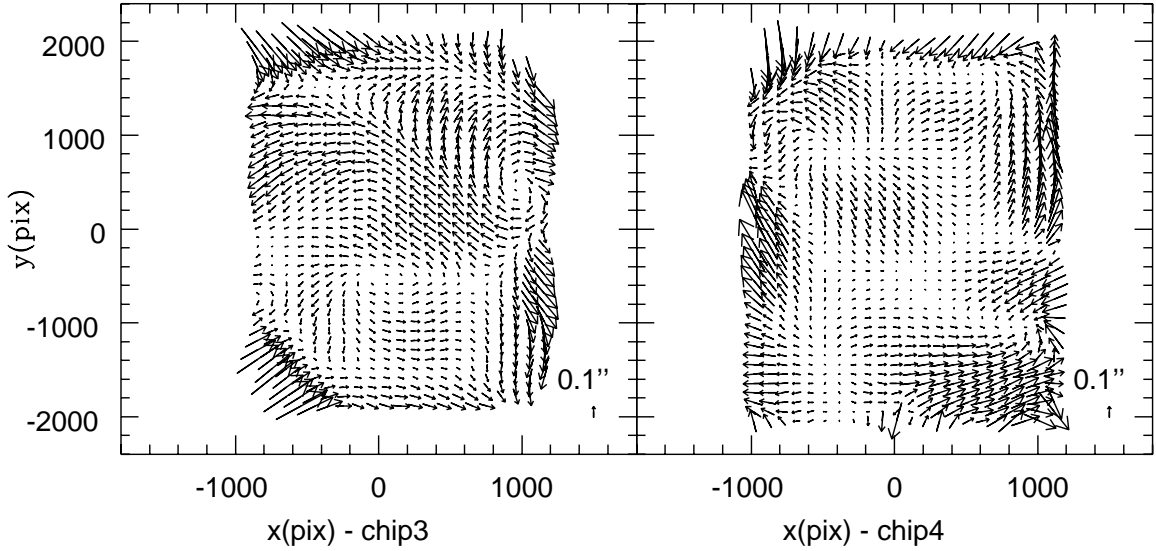}
\caption{Stacked position residuals as a function of the 
chip coordinates (x,y) after radial correction.
The $0\farcs1$ size vector is plotted on the lower right corner of the panels.
Only a linear characterization of the detector is calculated and applied.  
The fixed pattern shows that an additional correction is still needed,
and that the correction is different for each CCD. 
Similar plots for Chip~1 and Chip~2 present different distortions.}
\label{fig:distorsion}
\end{figure*}

Detectors with field curvature
introduce geometric distortion into the images.
The image will experience magnification to a factor
that depends on the distance to the optical axis.  
The resulting radial displacement is usually proportional to $r^3$, 
although perfect azimuthal symmetry is not always seen.
This effect must be corrected prior to any celestial coordinate calculation.

%This is the case for the INT-WFC. 
It is often found in the literature that
the radial distortion of the INT-WFC
is removed by following the instructions
given in the instrument web page.
It suggests performing a radial correction with origin in the optical center
of the mosaic (i.e., the composition of the four chips) 
to calculate a polynomial solution afterwards.

We first transformed the instrumental (x,y) coordinates of
all targets on all CCD into
global mosaic system $\rm (x_m,y_m)$ coordinates
with origin in the optical center, using
the linear equations:
\begin{center}
\begin{equation}
\left\{
  \begin{array}{l}
  \displaystyle x_m = a*x+b*y+c \\
  \displaystyle y_m = d*x+e*y+f
\end{array} \right.
\label{eq:sol}
\end{equation}
\end{center}
where the $a\, b\, c\, d\, e\, f$ coefficients for each CCD 
were taken from the 
technical reports
on the WFC.

The radial correction to the coordinates $\rm (x_m,y_m)$ is defined as follows:
\begin{equation}
  \Delta x = R\ x_m\ (x_m^2+y_m^2)\\
  \Delta y = R\ y_m\ (x_m^2+y_m^2)
\end{equation}
where R=$\rm -5.30\times10^{-10} \, pixel^{-2}$ is the distortion 
coefficient.

After the radial correction was applied, 
we used a general cubic polynomial solution
to describe the detector, as explained in the previous Section. 
We chose a third-degree polynomial so additional
distortions could be accounted for.

\begin{figure}[b!]
\centering
\includegraphics[width=0.45\textwidth]{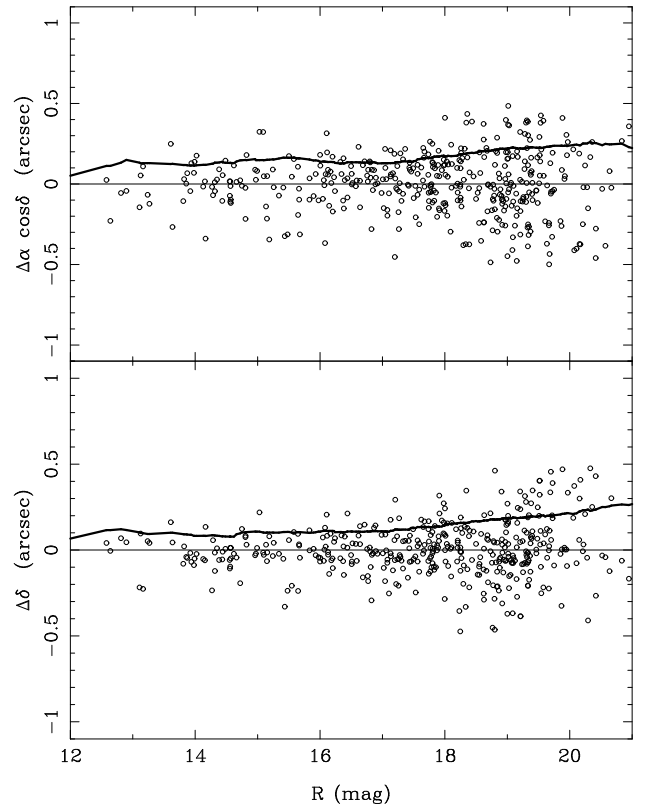}
\caption{
Difference between our catalogue and 2MASS positions
for the reference stars, as a function of R-magnitude.
Upper panel: difference in right ascension.
Lower panel: declination differences.
The dark line represents the standard deviation.}
\label{fig:mag}
\end{figure}

The residuals of the reference stars are the difference bet\-ween 
the transformed instrumental positions and the reference catalogue coordinates.
In Figure~\ref{fig:radial} we have 
plotted the vector field of residuals as a function of position 
in the focal plane. A fixed pattern
geometric distortion is still present in the INT-WFC system.
After several failed tests, we concluded that the
combination of a global mosaic radial correction and
a subsequent general polynomial solution cannot 
fully correct
the distortion caused by this telescope-camera combination.
%Figure \ref{fig:radial} shows the remaining residuals.
%The standard deviation of these differences 
%is around 1$\farcs$5 in each coordinate.

The reason why our characterization failed
is that each chip demands a different correction.
Figure~\ref{fig:distorsion} shows that after correcting
for radial distortion
systematic errors remain
in the relative positions (as expected)
but each CCD presents a very different 
pattern of distortion.
Hence, % in order to get the most accurate astrometric calibration
for a more accurate calibration
we deviated from the generally adopted strategy:
after correcting for  radial distortion,
we estimated an individual fourth-degree polynomial solution for each chip
as opposed to calcula\-ting a global solution for the whole mosaic system.
%We have chosen a fourth-degree polynomial in our final
%solutions to better compensate for distortions.

\subsection{Results and Error Estimate}

We estimate the uncertainty of our calibration
by comparing our final catalogue with a high precision one:
2MASS.
%By comparing our final catalogue with a high precision
%one, we may estimate the uncertainty of our calibration.
%The 2MASS catalogue was used as reference.
We plot the difference between our positions and
2MASS' in Figure~\ref{fig:mag}, as a function
of the R-magnitude of the targets. 
%As expected, the average of the difference is zero.
The difference between both catalogues is larger
in fainter objects but the average is zero, as expected.
The standard deviation of the difference is a more meaningful
parameter. 
If the complete catalogue is considered,
the standard deviation is 
($\sigma_{\alpha cos\delta}, \sigma_{\delta})$ = (0$\farcs$19, 0$\farcs$16).
Note that the standard deviation is larger (0$\farcs$5)
when a global solution for the whole mosaic is calculated.
%therefore we have managed to improve the astrometric accuracy.
The number of reference stars in each CCD is 140, 123, 138 and 206 stars, respectively.

Let us also remark the lack of any magnitude trend of the
differences: the standard deviation increases for faint objects,
but the average is always $\sim$0.
This fact evidences that the positions obtained are free 
of any obvious ``magnitude equation'',
%(assuming that 2MASS astrometric accuracy does not depend on magnitude).
%Magnitude equation is 
the undesired correlation between the target position
relative to the image center and the magnitude of the target.
We also investigated the dependence of position difference 
\textit{versus} color, and
no systematic difference was found.

%\section{Astrometric Catalog}
%\label{s:AST}
%
%\subsection{Comparison of the astrometric results}
%    Check with other catalogs (2MASS,Hipparcos). Show our improved astrometric solution.
%
%    Belen has found differences with 2MASS of up to 0''.2 at worst for V=21
%(smaller magnitude stars have smaller astrometric error). That's not the astrometry
%error, that is:  $\rm 0''.2 = \sqrt(\sigma^{2}_{SDSS}+\sigma^{2}_{ASTRO})$.
%There are no catalogs that reach as faint stars as ours and we cannot perform
%a comparison. $\rm \sigma^{2}_{ASTRO}$ represents the internal error,
%and can be quantified if the observations are repeated with the same instrument
%and telescope. 

%__________________________________________________________________

\section{AUTOMATIC SEARCH OF OB ASSOCIATIONS}
\label{s:code}

%OB associations are groups of young 
%stars, often gravitationally unbound,
%identified by the clustering of bright O and B stars \citep{Gal99}.
%They can be found in all Local Group galaxies
%with current star formation \citep{G02}.
%Typical sizes are $\la$ 100pc, but they can be part of
%larger scale structures. 

Catalogues of extragalactic OB associations meant as clustering of bright blue stars
over a haze of field stars
can already be found in the first extragalactic photographic jobs.
However, marking OB associations 'by eye' is a
subjective procedure; the results, namely the sizes of the found associations,
depended on a number of factors: distance to the galaxy, plate scale,
%(spatial resolution of the photographic plates)
exposure time and selection
criteria (see discussion of \citet{H86}).
%There is the additional arbitrarity of deciding whether a star grouping was one association or two
%nearby smaller ones \citep{H78}.

The advent of CCDs combined to target extraction routines (such as DAOPHOT)
has given rise to the automatic search of OB associations. 
The algorithms usually follow two main philosophies: (i) search
for statistically significant peaks (over 2 or 3 $\sigma$) in maps of stellar
density \citep{Kal96,Gal03} or (ii) the \textit{Path Linkage Criterion Method}
\citep{B91} also known as \textit{friends-of-friends algorithm}.
%\citep{B91,W91,BD92,I96,Gal99}.
The use of pattern recognition \citep{FB84} is less common.

We have developed an algorithm
based on the friends-of-friends method:
 two stars belong to the same
association if the distance of one star to the next
is less than a given amount.
The input entries to the code are
a list of points with spatial coordinates
and  two free parameters,
the search distance (\ds) and 
the  minimum number of members (\nm)
for a group to be considered an association.
%that an association must have to be considered so.

Given an OB star, the programme looks for other OB targets within the search 
distance and registers them as members of the same association. 
The search is
repeated for all the new OB members until no star is found within \ds~
from any of the peripheral stars.
In the process, non-OB stars are also included in the association
if they are less than \ds~ away from an OB member.
Groups with less than %\textbf{ 
\nm~ OB members %} 
are discarded.

The one case the algorithm would not detect
is a chain of stars each closer to the next less than \ds~
forming a ring of radius larger than \ds.
The code would walk the ring and miss the central targets.
At the moment, this has to be checked visually.
%The second one is that non-OB stars that fall between two 
%different OB associations may be included in both associations.

There is no physical constraint for
the parameters \ds~ and \nm.
In fact, friends-of-friends users do not agree on 
how to decide their values.
\nm~ is more problematic, since the expected population
of an association depends on the mass of its parent cloud and the IMF.
%In this work we will decide whether an OB association is valid from its
%color magnitude diagram, because it provides a more
%physical insight of the population.
%It is possible to use statistical tests to check against random clumping,
\citet{Bal04} decided \nm~ upon evaluating the chances that an association
is actually an accidental group of stars.
Other works consider that an OB association is valid if its stellar
surface density ($\Sigma$) is statistically above the mean,
and choose \ds~ and \nm~ on these grounds.
\citet{I96} uses \ds~ that maximizes 
the mean normalized fluctuation of $\Sigma$.
\citet{W91} sets $\rm D_S = 1 / \sqrt( \pi \times \Sigma)$
and \nm~ by comparison of the resulting associations to a random area
of the galaxy. \citet{Gal99} find \ds~ and \nm~ 
from the Fourier transform of the surface stellar density.

In this work we infer \nm~ from Milky Way associations,
by counting the members observable if they were located
at the distance of the studied galaxy 
%\textbf{
(see Sect.~\ref{ss:parms}). %}
For \ds, we follow
\citet{B91} and \citet{BD92} and set this parameter
to the abscissa of the maximum of the function $f_{N_{min}}(D_{S})$,
the number of associations that the algorithm finds
with at least \nm~ stars when run with search distance \ds.
At larger \ds~ associations tend to merge into a single larger system. 
On the other hand, if \ds~ is small then the associations
separate into their individual components (isolated
stars or binaries).
%Visual inspection of
%the associations found with our code for increasing
%\ds~ illustrates this effect very clearly. % (see Fig.~\ref{F:OBdistr}).
%This criterion has been followed by subsequent authors
%(e.g., \citet{Bal96,Bal98,Pal01,Bal04}) and
We adopt this criterion because it is %\textbf{
reproducible, %}
but we stress that it does not guarantee that the groups found in
this way correspond to physical associations.
%Nevertheless, lacking a physical
%criterion, an objective one is the best approach.
Note that the visual identifications of OB associations
already encountered this problem, and deciding whether a star clustering
is only one association, or it can be split into two or more smaller ones
was quite arbitrary \citep{H78}.
\citet{H86} already pointed out that additional information is required to make sure
that we are dealing with physical groups.
In fact, the same problem is found when defining 
OB associations in the Milky Way.

%\textbf{
\citet{Bal07,BGEG09} have recently shown
that there is a correlation of the typical size of the associations 
found by friends-of-friends algorithms and the \ds~ value used.
These results stress the fact that the OB associations identified
this way are not necessarily bound physical entities. Nonetheless,
enclosing OB stars in groups is a useful and necessary first step to study 
the youngest galactic population. Further work studying the color-magnitude
diagrams of the associations found in this work (Garcia et al., in prep.) 
will allow us to progress
towards the identification of the real associations.
%}

\section{OB ASSOCIATIONS IN IC~1613}
\label{s:run}

The catalogue of OB associations was made in
3 steps. We firstly built 
an input catalogue of candidate OB stars (Sect.~\ref{ss:input}).
Values for the two free parameters of the code, \ds~ 
and \nm, were decided afterwards (Sect.~\ref{ss:parms}).
Finally, the code was run to produce the list of associations,
which we inspected individually (Sect.~\ref{ss:val}).

\subsection{Input Catalogue of Stars}
\label{ss:input}

\begin{table}
\caption{Photometric-quality requirements for stars
to enter the algorithm.}
\label{T:phot}      
\centering                          % used for centering table
\begin{tabular}{c  c  c  c  r  c }        % centered columns (4 columns)
\hline\hline                 % inserts double horizontal lines
V-mag & \multicolumn{2}{c}{$\rm \#$-parm.} & \multicolumn{2}{c}{$\rm \chi$-parm.} & error(U, B, V) \\    % table heading 
      & MIN         &      MAX             & MIN         &      MAX              &    MAX     \\  
\hline                        
   $\le$ 19.0 & -2 & 2 & 0.5 & 30 & 0.05 \\   
   $ > $ 19.0 & -2 & 2 & 0.5 &  4 & 0.05 \\   
\hline                        
\end{tabular}
\end{table}

The starting point is the deep 
photometric catalogue described in Sect.~\ref{s:OBS}.
We kept only stars with exceptional photometric quality,
setting hard constraints on the errors of  U, B and V
magnitudes. %: 0.05 at worst.
The reason is that we built the list of likely OB stars in IC~1613 
based only on their reddening-free Q color (see below),
a linear combination of U, B and V.
If allowed errors in these magnitudes
were as large as 0.1, the resulting Q error would be 0.2
and could lead to target mis-classification.
We required errors smaller
than 0.05 magnitudes.
Note that by setting hard constraints on UBV, we 
do not include the red population
with faint U magnitudes (and large U-errors).

We set looser constraints on
the DAOPHOT parameters that describe the goodness of the PSF extraction of each target.
We allowed for a \#-parm. interval (roundness of the PSF) of $\rm [-2,2]$.
%\textbf{ 
The $\chi$ parameter (quality of the PSF fit, hereafter $\chi$-parm.) %}
increases with magnitude and tight constraints
on this parameter may leave out the brightest stars.
In fact, two trends of the $\chi$-parm. are seen in Fig.~\ref{F:phot},
corresponding to the long and
the short exposures. 
DSS, GALEX and/or VIMOS images 
reveal that most of the faint targets
with $\rm \chi > 5$ are background galaxies or blends.
A very small fraction corresponds to stars that saturate.
We will allow for $\rm \chi \leq 4 $ for faint stars,
and  $\rm \chi \leq 30 $ for bright targets.

The photometric qualification criteria
are summarised in Table~\ref{T:phot}. 
Fig.~\ref{F:phot} shows the typical photometric errors of
the complete catalogue (black), with 
the targets that comply with the photometric requirements marked in grey.
From a total of 84,771 targets
in the initial catalogue, we keep 3,960.

\subsubsection{Candidate OB Stars}
%\subsubsection{Finding Blue Massive Stars from their Colors}
\label{sss:Qchart}

\begin{figure}
  \centering
  \includegraphics[width=9cm]{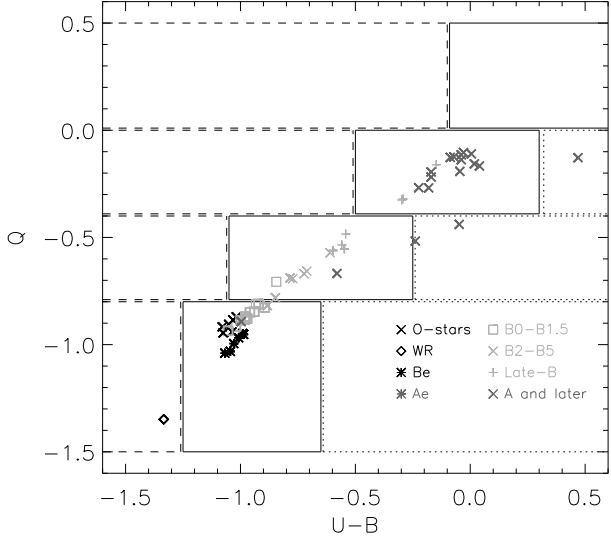}
  \caption{Locus of O and B stars in the \ubq~diagram.
   For each Q-interval delimited by a horizontal band,
   the central box corresponds to unreddened
   or moderately reddened objects.
   The symbols represent stars of different spectral type
   (from \citet{Fal07}) as detailed in the legend. 
   Most O stars and B supergiants
   are found in the central unreddened box of the two bottom strips.
  \label{F:Q}}
\end{figure}

Candidate blue massive stars were chosen from their 
Q color, the \textit{reddening-free} parameter
$ Q = (U-B) - 0.72 \times (B-V)$.
Q increases monotonically towards
later spectral types in the interval $\rm Q \in [-1.0,-0.4] $, corresponding to
%\textbf{
O3-A0 supergiants and O3-B5 dwarfs
%}
(see for instance the calibration of \citet{F70}).
We defined the locus of O and B stars in the \ubq~ diagram (Fig.~\ref{F:Q})
using the colors of stars with spectral types determined
by \citet{Fal07}.
While Q is an indicator of spectral type,
U-B holds the information of whether the star is reddened or not. 
Most O and B supergiants are found in unreddened boxes with $\rm Q \leq -0.4$.
IC~1613's known WO is found in 
the diagram as an outlier in the lower-left corner.

Q was the only selection criteria to find OB stars.
We allowed for wide ranges of U-B and B-V 
to account for reddening
since we suspect that IC~1613's internal extinction is not negligible,
particularly for OB associations,
and that the dust content may be significant in the bar
(see full discussion in Sect.~\ref{sss:ext}).

The allowed V interval is $\rm V \in [16,24]$.
This range ensures that all the interesting targets for our spectroscopic studies
are included:
%The brightest/faintest blue massive stars we expect to see at the distance of IC~1613 have 
O3~I/B0~V types, with $\rm M_V$=-6.5/-3.8 
\citep{Massey98}. At the adopted distance modulus DM=24.27,
the observed  magnitudes will be V=17.8/20.5. We allow for extra 1.5 magnitudes
in the bright interval limit, not to leave
out exceptional cases such as hypergiants. The limit for faint
stars is also extended to allow for reddening and include 
dwarfs of later B-types. 

Table~\ref{T:OB} summarizes  
the photometric properties of candidate OB stars.
The photometric constraints set on non-OB stars
are listed in Table~\ref{T:other}.
%; the only difference is the allowed range of Q.
This is not an attempt to determine the
exact spectral sub-types of OB stars from Q.
The degeneracy of blue massive star colors
at optical wavelengths may be
broken in the future with improved photometric capabilities
combined with new statistical techniques \citep{M-A04},
but the required accuracy  (one thousandth of a magnitude)
is not feasible to date.

\begin{table}
\caption{Photometric constraints for OB stars}
\label{T:OB}      
\centering                          % used for centering table
\begin{tabular}{c | c | c }        % centered columns (4 columns)
\hline\hline                 % inserts double horizontal lines
Parameter & MIN & MAX \\    % table heading 
\hline                        
V          &   16 &   24 \\
Q-parm     & -2.0 & -0.4 \\
U-B        & -2.0 &  2.0 \\
B-V        & -0.5 &  2.0 \\
%error V    & 0.05 & 0.05 \\
%error B    & 0.05 & 0.05 \\
%error U    & 0.05 & 0.05 \\
\hline                        
\end{tabular}
\end{table}

\begin{table}
\caption{Photometric constraints for non-OB stars}
\label{T:other}      
\centering                          % used for centering table
\begin{tabular}{c | c | c }        % centered columns (4 columns)
\hline\hline                 % inserts double horizontal lines
Parameter & MIN & MAX \\    % table heading 
\hline                        
V          &   16 &   24 \\
Q-parm     & -0.4 &  1.0 \\
U-B        & -2.0 &  2.0 \\
B-V        & -0.5 &  2.0 \\
\hline                        
\end{tabular}
\end{table}

%\textbf{
We note here that the use of the Q parameter has two caveats.
(i) It is reddening-independent as long as the reddening law is standard.
(ii)
Beyond very early spectral types the univocal behaviour of Q
cannot be ensured.
At Q$\sim$-0.4,  it varies rapidly with spectral type and luminosity class.
%while Q=-0.48 for B9I, A0I has Q=-0.45 and  A2I Q=-0.32  ;
%      Q=-0.50 for B4V, B5V has Q=-0.46 and  B6V Q=-0.39
%(see \citet{F70}'s database).
The inherent scatter of the calibration of Q with spectral type
%plus small or moderate photometric errors 
may take mid/late B dwarfs and late-B/early-A supergiants to values larger than -0.4. 
This is the case for the three ``late-B'' stars in the Q$>$-0.4 box of Fig.~\ref{F:Q}.
%that are actually \citet{Fal07}'s  A4 (B9 II, Q=-0.33), B1 (B9 II, Q=-0.16)
%and B18 (B9 II, Q=-0.32).
%The case of B1 seems a bit more extreme, but examining
%its spectrum the CaII~K/(CaII~H+H$\rm_{\epsilon}$) ratio
%is intermediate between a B9 and an A0 supergiant
%which, together with the steep slope of the Q function 
%in this range, is consistent with a redder Q color.
%Similarly, late-A to early-G type dwarfs (no longer in the univocal regime of Q)
%have  Q greater than -0.4 but very similar,
%and thus may contaminate the list of candidate OB stars,
%as it happened with a foreground G2 and a halo AV star
%that entered the Q$<$-0.4 box of Fig.~\ref{F:Q} as ``A and later'' points.
Late stars around G-type are close to a local minimum of the
Q \textit{vs} spectral type relationship, at Q$\sim$-0.37.
The scatter of the calibration or specific stellar properties (UV excess
from stellar activity, for instance) may bring the stars to the boxes
of the early types.
Finally, \citet{Fal07}'s A8 (A2~Ia) is also in the Q$<$-0.4 locus of B stars.
The authors found $\rm H_{\alpha}$ and $\rm H_{\beta}$ P~Cygni profiles,
and their published spectra shows that the Balmer lines in the blue
and near-UV are unusually weak.
Clearly, this star is peculiar and hence not expected to follow
the general trend.
%}

\begin{table*}
\caption{Brightest candidate OB-stars: magnitude and colors. 
Column 'ASSOC' indicates
the association the stars belong to as found in this work 
(-1 means they are isolated). We also provide identification
numers and spectral types from  \citet{Fal07}, when available.}
\label{T:ttbright}      
\centering          
\begin{tabular}{c c c c c c c c r r}
\hline\hline  
ID & RA$\rm [deg]$& DEC$\rm [deg]$& V  & B-V & U-B & Q & ASSOC & \multicolumn{2}{c}{\citet{Fal07}}\\
   &(J2000.0)&(J2000.0)&    &     &     &   &       & ID & SpType\\
\hline
39916  &  16.194780  &  2.166142 &  16.43  &   0.54  &  -0.05  &  -0.44  &    -1  & C12 & G2 \\
59802  &  16.246347  &  2.154164 &  16.44  &   0.12  &  -0.58  &  -0.67  &   127  & A8  & A2~Ia \\
83197  &  16.369705  &  2.083662 &  16.67  &   0.19  &  -0.51  &  -0.64  &    -1  & -   & - \\
52033  &  16.224657  &  2.119503 &  17.24  &   0.63  &  -0.02  &  -0.47  &    -1  & -   & - \\
62390  &  16.253813  &  2.178189 &  17.42  &  -0.14  &  -0.98  &  -0.88  &   137  & A10 & B1~Ia \\
69336  &  16.276543  &  2.158706 &  17.69  &  -0.15  &  -0.98  &  -0.87  &   185  & B3  & B0~Ia \\
15779  &  16.124454  &  2.051565 &  17.76  &   0.64  &  -0.10  &  -0.56  &    -1  & -   & - \\
42294  &  16.200409  &  2.079537 &  17.95  &   0.65  &   0.01  &  -0.46  &    60  & -   & - \\
65875  &  16.264278  &  2.162671 &  18.06  &   0.01  &  -0.55  &  -0.55  &   161  & A11 & B9~Ia \\
 1705  &  16.023046  &  2.209241 &  18.13  &   0.44  &  -0.16  &  -0.48  &    -1  & -   & - \\
60449  &  16.248133  &  2.154477 &  18.23  &  -0.13  &  -0.97  &  -0.88  &   127  & B4  & B1.5~Ia \\
56665  &  16.237545  &  2.084683 &  18.24  &   0.37  &  -0.41  &  -0.68  &   114  & -   & - \\
23263  &  16.151031  &  2.190748 &  18.24  &   0.39  &  -0.24  &  -0.52  &    -1  & C5  & A~V \\
58040  &  16.241360  &  2.096872 &  18.39  &  -0.09  &  -0.88  &  -0.82  &   120  & A7  & B2~Iab \\
27381  &  16.163216  &  2.146994 &  18.42  &  -0.07  &  -0.88  &  -0.83  &    -1  & -   & - \\
68599  &  16.273815  &  2.096740 &  18.54  &  -0.09  &  -0.85  &  -0.78  &    -1  & B13 & B5~Iab \\
65133  &  16.261927  &  2.083119 &  18.56  &  -0.05  &  -0.70  &  -0.66  &   163  & -   & - \\
52841  &  16.226801  &  2.136780 &  18.62  &  -0.07  &  -0.72  &  -0.67  &   105  & A5  & B3~Ib \\
34964  &  16.182615  &  2.112548 &  18.62  &  -0.13  &  -1.04  &  -0.94  &    43  & B11 & O9.5~I \\
18570  &  16.135248  &  2.158957 &  18.68  &  -0.13  &  -0.94  &  -0.85  &    21  & C6  & B1~Ia \\
%57798  &  16.240717  &  2.097148 &  18.71  &   0.62  &  -0.29  &  -0.74  &   120  \\
%67841  &  16.270922  &  2.142758 &  18.71  &  -0.15  &  -0.96  &  -0.85  &   179  \\
%57628  &  16.240234  &  2.100318 &  18.71  &  -0.13  &  -0.89  &  -0.80  &    -1  \\
%57613  &  16.240185  &  2.078899 &  18.78  &  -0.09  &  -0.89  &  -0.83  &   122  \\
%35071  &  16.182901  &  2.084464 &  18.78  &   0.65  &  -0.52  &  -0.99  &    -1  \\
%79536  &  16.334723  &  2.126752 &  18.80  &   0.50  &  -0.19  &  -0.55  &    -1  \\
%27882  &  16.164688  &  2.194739 &  18.80  &   0.50  &  -0.13  &  -0.49  &    -1  \\
%34231  &  16.180744  &  2.172854 &  18.83  &  -0.17  &  -0.93  &  -0.81  &    -1  \\
%72900  &  16.292332  &  2.208207 &  18.83  &  -0.19  &  -0.84  &  -0.71  &    -1  \\
%64227  &  16.259102  &  2.143061 &  18.84  &  -0.14  &  -1.02  &  -0.92  &   151  \\
%53145  &  16.227625  &  2.141626 &  18.87  &  -0.14  &  -1.00  &  -0.89  &    -1  \\
%69476  &  16.277063  &  2.158948 &  18.89  &  -0.22  &  -1.13  &  -0.97  &   185  \\
%45453  &  16.208094  &  2.117661 &  18.92  &  -0.05  &  -0.60  &  -0.56  &    69  \\
%52635  &  16.226233  &  2.100556 &  18.96  &  -0.08  &  -0.71  &  -0.66  &   106  \\
%69217  &  16.276066  &  2.178654 &  18.96  &  -0.22  &  -1.08  &  -0.92  &   187  \\
%63932  &  16.258221  &  2.134749 &  18.97  &  -0.21  &  -1.03  &  -0.88  &   151  \\
%69239  &  16.276150  &  2.157813 &  18.98  &  -0.23  &  -0.97  &  -0.81  &   185  \\
%64029  &  16.258501  &  2.173535 &  19.00  &   0.62  &  -0.19  &  -0.64  &   158  \\
\hline 
\end{tabular}
\end{table*}
                %\label{T:ttbright}      

We find that the candidate OB stars concentrate in Chip~4,
mostly on the NE lobe of the galaxy where the 
giant bubbles are seen, 
%\textbf{ 
but also in the opposite direction,
in the SW side of the giant \ion{H}{i} cavity where
part of the galaxy is enclosed. %}
This is in agreement with previous results (see Sect.~\ref{s:intro}).
%\citet{EE80} noted higher concentration of massive
%stars at the ends of the bars of  magellanic irregulars
%and suggested that they may experience gas dynamics similar to those in 
%the inner barred regions of more massive spirals.
Despite our large allowance for V, U-B and B-V
to include the most reddened targets, 
we find very few blue stars in the bar.

%We find most of massive blue star candidates within chip~4,
%as \citet{BAG07} pointed out.
%BAG07   Most of main  sequence stars (age <= 500Myr) inside a 10' radius, and fit chip4
%They concentrate mostly around the ends of the bar.
%\citet{EE80} noted this fact for  magellanic irregulars
%and suggested that they may experience gas dynamics similar to those in 
%the inner barred regions of more massive spirals.
%As \citet{Cal99}, we also find that young blue stars avoid 
%the very center of the galaxy.

The 20 brightest blue stars from our catalogue of OB candidates are listed
in Table~\ref{T:ttbright}. 
Spectral types are known for 12 of them.
This subset includes an A dwarf and a G2 star
%\textbf{
(the two ``A and later" stars with Q$\rm \lesssim$-0.4 in Fig.~\ref{F:Q}), 
which are foreground objects according to \citet{Fal07}. %}
It is remarkable that only one red star has entered Table~\ref{T:ttbright},
supporting our color selection criteria.
Despite its red intrinsic colors
the G2 star is in the limit of the accepted Q interval,
justifying the need of
hard constraints on the photometric errors of stars.
The remaining 10 targets (3 of them brighter than V$<$17.8)
are O, B and A supergiants, with high chances
of been IC~1613 members.
Table~\ref{T:ttbright} supports our selection method
and also indicates that the foreground contamination is small.

\subsubsection{Variable Extinction and Extinction Map}
\label{sss:ext}

The foreground extinction towards IC~1613 is remarkably low 
%\textbf{
(\ebv $\sim$0.2, see Sect.~\ref{s:intro}). %}
In the past it was argued that the internal reddening was also negligible
since background galaxies can be seen 
through IC~1613 \citep{H78,F88,KH00}, although \citet{H78} 
already marked the position of $\sim$10 possible dust clouds in the galaxy.

More recent papers yield higher values of extinction in the lines of sight of
individual objects. 
\citet{Hu80} 
%assigned unreddened colors to 
%stars of known spectral types,
derived  \ebv=0.05-0.10 towards early type members,
and $\sim$0.2 towards M-supergiants.
The cepheid studies of the OGLE \citep{WB01}
and Araucaria \citep{Pal06} projects yield average extinction values
towards their targets of
$<$\ebv$>$= 0.085 $\pm$0.052 and 0.090$\pm$0.019,
with extinction law similar to the MW.
\citet{Tal07}'s  study of  three M-supergiants yields
three different values \ebv=0.04, 0.06 and 0.03.
In the bubble region,
\citet{Gal99} report \ebv=0.06
and \citet{Lal02} finds \ebv~ in the range of 0.18-0.4 towards 9 OBA stars,
considerably larger.
The extinction towards the WO is 0.1 \citep{Lal03}, but
reddening is suspected to be high and anomalous towards Wolf-Rayets.
Detections of the molecular cloud, neutral hydrogen
and dust by IRAS and Spitzer in the bubble region
\citep{MCW88,Jal06} are further evidence of extinction in the area.

   \begin{figure}[b!]
   \hspace{-0.5cm}
%   \centering
   \includegraphics[angle=0,width=0.5\textwidth]{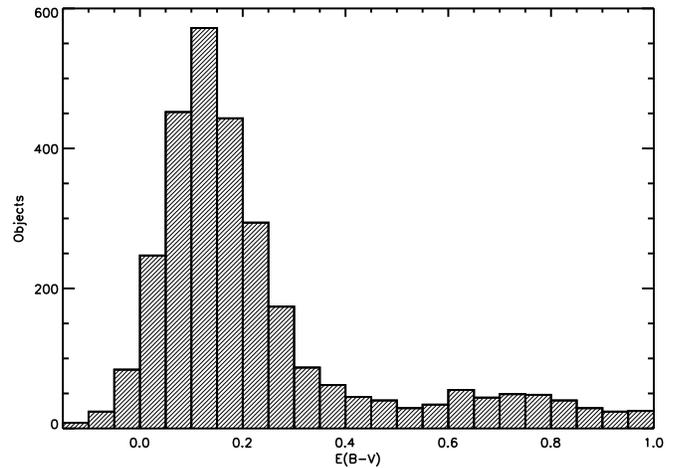}
      \caption{Extinction distribution towards the blue massive star candidates of IC~1613.
       Foreground reddening has not been corrected.
       Most of the stars have \ebv~ in the range 0.05-0.25 and
       the peak of the distribution
       is located at 0.10-0.15, significantly higher than the 
       published foreground extinction.
       }
       \label{F:extH}
   \end{figure}

   \begin{figure*}[htpb!]
%   \centering
   \hspace{-1.cm}
   \includegraphics[angle=90,width=20cm]{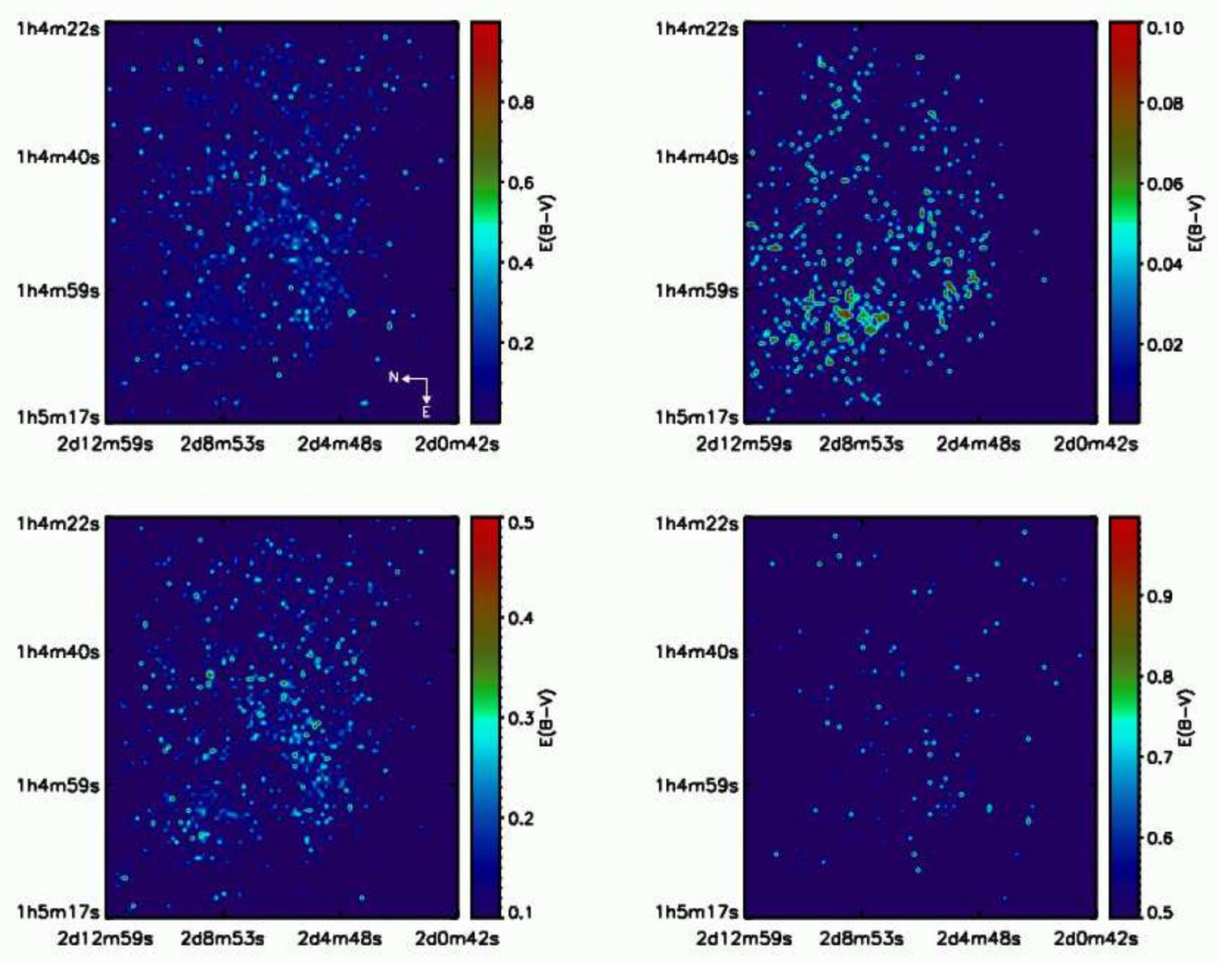}
      \caption{Extinction maps.
Only the portion of IC~1613 containing OB associations is shown
(same as in Fig.~\ref{F:all}).
       Different panels consider different \ebv~ ranges. Clock-wise starting from top-left: (1) complete range 0.-1.0,
       (2) the traditionally considered range for this galaxy 0.-0.1, (3) intermediate \ebv~ range 0.1-0.5,
       and (4) large extinction 0.5-1.0.
       }
       \label{F:extM}
   \end{figure*}
%   \clearpage

We therefore suspect that internal reddening 
is highly variable and far from negligible.
To prove it, we have estimated the color excess in the lines
of sight of our candidate OB stars in IC~1613.
Intrinsic colors were calculated from Q,
thanks to its univocal relation with spectral type
(hence $\rm (B-V)_0$) that holds for the early-type stars.
%Derivation of \ebv~ is then straightforward by subtraction
%to the observed colors.
This relation was parameterized for low-metallicity stars
by \citet{MWD00}:
\begin{equation}
\label{eq:Qmass}
(B-V)_{0} = -0.005 + 0.317 \times Q
\end{equation}
using Kurucz' ATLAS9 Z=0.08Z$_{\odot}$ models.
(No such relationship exists for metallicity as low as
IC~1613's, as far as we know).
This is an approximation, since equation~\ref{eq:Qmass} does not hold
%This is an approximation, since
%stars will not follow this function
if the reddening law deviates from the standard behaviour.
%\textbf{
Another limitation of equation~\ref{eq:Qmass}
is that it is independent of luminosity class.
Although convenient for fast calculations, 
it will yield $\rm (B-V)_0$
slightly bluer than the actual intrinsic colors of supergiants and
slightly redder for dwarfs. The effect will have a maximum for B-type stars.
For instance, the comparison of the \ebv~ provided by \citet{Fal07} for seven early-B
supergiants (0.06 on average) with the values derived from equation~\ref{eq:Qmass}
(0.15 on average), hints that the latter overestimates extinction for these stars.
However, we
expect that statistically these effects will compensate in the large stellar
population of the galaxy.
%}

The distribution of the extinction towards the blue objects of IC~1613
is shown in Fig.~\ref{F:extH}. It peaks in the \ebv~ interval 0.10-0.15.
Most of the stars have color excess ranging 0.05 to 0.25,
higher than the foreground extinction provided in the literature.
For a few objects we obtain \ebv$>$1.
Most of them (more than 90\%) are background galaxies for which
equation~\ref{eq:Qmass} does not apply, and have been discarded.
%It is puzzling that some of these galaxies can be seen through
%the bar, and it can be considered further proof of inhomogeneities
%of the reddening distribution.
%\textbf{
The negative values of \ebv~ in Fig.~\ref{F:extH}
are due to a combination of factors, mainly
the intrinsic scatter of the calibration of the Q parameter
with $\rm (B-V)_0$, but also local changes in the slope of the
reddening law or photometric errors.
%}

The spatial distribution of \ebv~ is shown in Fig.~\ref{F:extM}.
The different panels are the extinction maps for 
4 different extinction intervals: the complete range (0.-1.0),
the traditionally considered range for this galaxy (0.-0.1), intermediate interval (0.1-0.5),
and large extinction (0.5-1.0).
%Objects with \ebv$>$1.0 are not shown since they are mostly galaxies.
Starting from the pixel map of
the INT-WFC Chip~4 image,
stars were distributed into boxes of $\rm 25 \times 25$ pixels.
Each box was assigned the average extinction
of the enclosed stars with \ebv~ in  the studied range.
%The \ebv~ values of all stars falling in a given box were considered as long
%as it was included in the studied \ebv~ interval. The average \ebv~ value was
%assigned to the enclosing box.
%Each box was assigned the average EBV~ of the contained stars with EBV~ values in the considered range.
Empty boxes (i.e., bins with no stars with \ebv~ in the considered interval)
where assigned the minimum of the interval.
In the 0.-1.0 and 0.-0.1 panels, empty boxes are assigned the
minimum 
%\textbf{
positive %} 
value found in the sample \ebv=0.0002.

Boxes with \ebv~ in the range 0.-0.1 trace well the star forming regions of the
galaxy, which is also the preferred location of OB star candidates.
A concentration of \ebv$\sim$0.1 points is seen at the bubbles
and at the SE edge of the galaxy.
Points with larger extinction  values (0.1$<$\ebv$<$0.5)
also trace the star forming regions, with a slight concentration towards
the bar area 
%\textbf{
and the SW part of the \ion{H}{i} cavity. %}
%are more disperse, but there is a slight concentration
%in the star forming regions. 
Larger \ebv~ values are distributed almost
randomly throughout the galaxy.
%, although the \ion{H}{i} bar is clearly traced.
These results support that extinction is not uniform, with rapid 
spatial variations.

\subsection{Free Parameters of the Code}
\label{ss:parms}

We followed the procedure described in Sect.~\ref{s:code}
to determine \ds~ and \nm.

We used  \citet{Hu78}'s catalogue of Galactic OB associations to
estimate \nm.
We removed all luminosity class-V stars with spectral type 
later than B5.
%\textbf{B5}. 
%\textbf{
( According to the V-magnitude limits of Table~\ref{T:OB}
the catalogue of OB candidates 
would include spectral types down to B8~V,
but the Q constraints set the cut at B5~V).
25 out of 78 associations (i.e., $\sim$30\%) are left with 3 or less OB members.
Since projection effects may be at work, 
%}
%16 out of 67 associations (i.e., 24\%) are left with 3 or less OB members.
%Since projection effects may be at work, 
%and eventhough  they are likely as significant
%or more in the Milky Way than in  IC1613,
%%
%%two stars in the Galaxy might look
%%separated by a given distance but in fact they are further
%%away on the depth-direction.
%%Anyway, this can be taken as an effect whose impact is statistically
%%the same in the Milky Way and 
%%IC1613, and we may consider the Galactic estimation for OBdist good.
we consider that 2 members
are too few for an extragalactic OB association. 
We therefore
set the minimum number of stars to 3,
at the risk of neglecting the less massive associations.
We expect that we will be able to spot out 
false OB associations through inspection of the images, and
the color-color and color-magnitude diagrams.
%On the other hand, our criteria will neglect the less massive
%OB associations. 

\begin{figure}
  \centering
  \includegraphics[width=9cm]{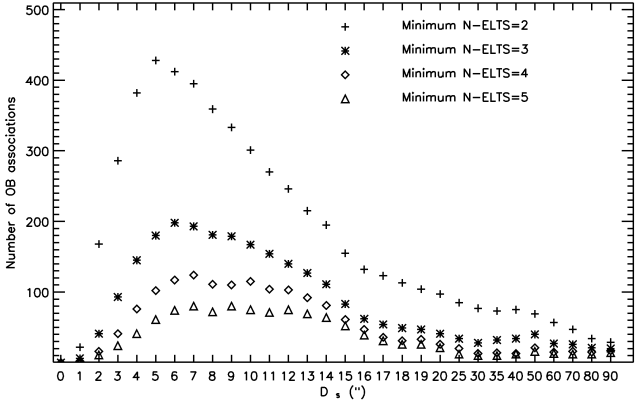}
  \caption{Number of OB associations found as a function of \ds~
   and \nm, $ f_{N_{min}}(D_{S})$. 
   Note the irregular scale on the x-axis.
   Different symbols represent the results for different values
   of \nm~ as indicated in the legend.
   There is a local maximum
   around \ds=6\arcsec. The slight increase at \ds=50\arcsec~ is caused
   by stars in the galactic outskirts, far away between them, that cluster
   because of the large \ds~ admitted value.
  \label{F:OBdist}}
\end{figure}

We ran the code for different search radii in order to
assess $f_{N_{min}}(D_{S})$ (see Fig.~\ref{F:OBdist}),
and set \ds~ to the abscissa of the maximum.
%Besides \nm=3, we tested three additional values.
%Independently of this parameter,
%there is a maximum for \ds~ at 6\arcsec.
%\textbf{
Besides \nm=3, we tested three additional values: \nm=4, 5 and 6.
The bump of $f_{N_{min}}(D_{S})$ flattens as \nm~ increases,
but, in the four cases there is a local maximum at \ds=6-7\arcsec.
This local maximum is preserved up to \nm=10 (not shown in Fig.~\ref{F:OBdist}).
%}

We therefore set
$\rm D_{S} = 6 \arcsec \; and \; N_{min} = 3 $.

%With this criteria,  an association of characteristics similiar to the Orion Nebula
%Cluster (maximum inter-star distance is roughly 0.05pc) would be included. 
%% Orion@450pc
%% Distancia mayor entre estrellas del trapezium  25'' (medido con Aladin sobre imh del 2mass)
%% Estan separadas como mucho 0.05pc

\ds=6\arcsec~ is similar to the values derived by Borissova et al. (IC~1613, 
6.6\arcsec) and Pietrzynski et al. (NGC 300, 7.7\arcsec), and much larger than those 
adopted by Bresolin et al. (1996, 1998, 0.83 to 3.9\arcsec~ in different spiral galaxies
at $\sim$10 Mpc). 
Since Borissova et al., Pietrzynski et al. and this work have used
2m-class ground-based telescopes with comparable instrumentation
whereas Bresolin et al. used HST-WFPC2 data,
it is tempting to interpret the differences
in terms of the superior
plate scale of the WFPC2 (0.1\arcsec/px) breaking the OB associations
into smaller pieces.
However, the physical values of \ds~ vary between
the 21 and 23 pc obtained by Borissova et al. and ourselves for IC 1613 and
the 35-84 pc obtained for spiral galaxies by Bresolin et al. and Pietrzynski et al.
Therefore, the smaller angular value of \ds~ obtained by Bresolin et al.
is a consequence of the longer distances towards their sample galaxies
(hence smaller  angular diameter of associations).
%Therefore, the smaller angular value of \ds~ obtained by Bresolin et al.
%does not imply that the superior
%plate scale of the WFPC2 (0.1\arcsec/px) can resolve the OB associations
%into smaller pieces, but instead that the angular diameter of the associations is
%smaller because of the distance towards their sample galaxies.
Whether the smaller physical values found for IC~1613 are due to
inherent differences of
star forming regions of irregular galaxies versus spirals,
the distance to the galaxies or to poor
statistics must be clarified in future works.
%\textbf{
In the light of \citet{Bal07}'s results we also anticipate
that the different \ds~ values used will result in different 
typical sizes for the associations, but we leave the discussion
on whether or not there exists a universal length-scale
of OB associations for a subsequent paper.
%}

% WFPC2 is 0.1''/px, and the central chip is 0.046''/px
%HST is a 2.4-meter reflecting telescope

\subsection{Final OB Association List}
\label{ss:val}

We ran our automatic search algorithm
on the catalogue of candidate OB stars one last time with 
$\rm D_{S} = 6 \arcsec \; and \; N_{min} = 3 $,
and produced the list of OB associations presented in Sect.~\ref{s:cat}.
All stellar groups provided by the code were accepted as valid associations,
as long as they have at least 3 valid members
(i.e., not background galaxies, blends or bright
red objects with Q$\sim$-0.4).
Each association was studied thoroughly using images, 
color-color (\cc~ and \ubq) and color-magnitude (\cmd~ and \qv) 
diagrams to discard fake detections.
The analysis of these diagrams will be presented in a second paper.

We examined high resolution images of IC~1613 from the 
HST archive and VLT-VIMOS
using the Aladin software \citep{aladin}
to discard non-valid association members, 
find neighboring associations and/or nearby objects
that did not enter the input catalogue because of the quality of their detection
(poorer than the requirements of Table~\ref{T:phot}).
%The examined images were from the HST (WFPC2) archive
%and our own INT-WFC and VLT-VIMOS data.
%We found that targets resolved by HST or VLT
%were sometimes blended in the INT images.
%In this case we trusted 
%DAOPHOT's PSF extraction, as long as each target had its own detection
%in our catalogue.
%This is not always the case: we have found several examples 
%for  nearby faint stars in V images,
%one of them clearly seen in the U-band and the other in the red-band,
%with only one photometric detection.

We also cross-matched our catalogue with published
spectral classification for IC~1613 targets.
The known spectral types of association members
are provided in Sect.~\ref{ss:comments}.

The code finds 198 groups. 
5 of these groups are galaxies.
After visual inspection, 
29 associations were discarded because
they did not reach the minimum number of valid members
%(see Sect.~\ref{s:cat})
and 47 were marked
dubious because we could not assure that they had at least \nm~
isolated blue stars.
The remaining 117 groups ($\sim$60\%) are bona-fide OB associations.

\section{CATALOGUE OF OB ASSOCIATIONS}
\label{s:cat}

    The list of OB associations built with our code
is presented in Table~\ref{T:ASSprop}.
Their positions are marked on the galaxy in Fig.~\ref{F:all}.
Zooms into different galactic sectors are provided in Appendix~\ref{S:f-charts}.

The associations with more than 10 OB members are listed
in Table~\ref{T:ttpopulated}, as illustration of the complete catalogue.
%A fraction of the catalogue is provided
%in Table~\ref{T:ttpopulated},
%where the associations with more than 10 OB members are listed.
%The columns of Table~\ref{T:ASSprop} are the following:
The content of different columns is:
\textbf{(1)} OB association identification number. \textbf{(2)} and \textbf{(3)} Right Ascension (RA)
and Declination (DEC) of the centre
determined from the average of the positions of the OB members, equinox J2000.0
\textbf{(4)} Number of OB members ($\rm N_{OB}$) of the association.
\textbf{(5)} Total number of members registered ($\rm N_{TOT}$); note that
our photometric selection criteria is biased against red objects
because we demand high accuracy in U magnitudes.
\textbf{(6)} Quadrant allocation in  Fig.~\ref{F:all}.
\textbf{(7)} Notes: 'Gal' -- the association is actually a galaxy; 'd' --
 association discarded because it does not have enough valid members
(i.e. visual inspection reveals
that some are blends or gas knots);
'?' -- dubious association (associations with a small number of members,
with a significant fraction under suspicion of being blends or not blue
stars);
'n' -- an entry in Sect.~\ref{ss:comments}.

   \begin{figure*}[tb]
%   \hspace{-3.5cm}
%   \includegraphics[angle=0,width=28cm]{./FIGURES/Assoc_all.ps}
   \centering
   \includegraphics[angle=180]{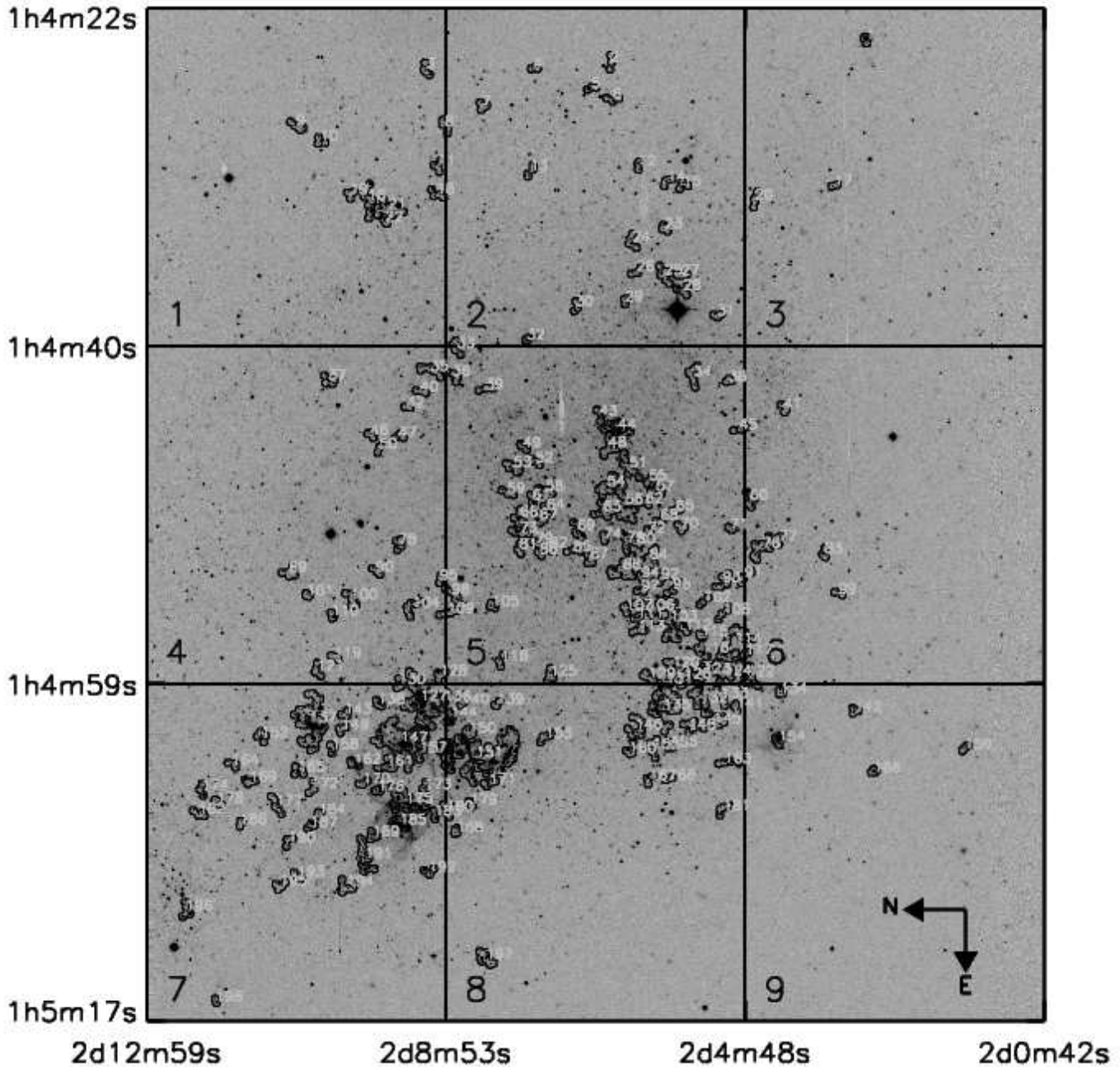}
      \caption{INT-WFC U-band image of IC~1613 (only the portion of Chip~4
       that contains the associations is shown). 
       The positions of the associations found in this work are marked in black.
        %%%%%%%%% PARA FIGURA EN COLOR
%       The RGB composition is made of the U- (blue), V- (green) and R-bands (red).
%       The positions of the associations found in this work are marked in red.
       The grid divides the image in 9 portions. A zoom into each quadrant
       is provided in Appendix~\ref{S:f-charts}.
       Note how the OB associations cluster in sectors 5, 7 and 8.}
%       around the eastern edge of the bar.}
       \label{F:all}
   \end{figure*}

%{\tiny
%\addtocounter{table}{5}       % table moved to after RFs
%\input{TABLES/T_OB}           % label {T:OBass}
%\addtocounter{table}{6}       % table moved to after RFs
%\input{TABLES/T_ASSOC_sum}   % label {T:ASSprop}
%}

The code also provides us with a list of the
OB members of each association and also the less massive population.
Defining OB associations by their members, rather than by hand-drawn
lines on photographic plates, allows a direct comparison of results between different
works and with observations.
% (if, for instance, we want
%to compare the distribution of OB associations with \ion{H}{i} maps).
The member listings will be available on-line;
an example is provided in Table~\ref{T:OBass}.

\begin{table}
\caption{OB associations with more than 10 OB-candidate members.}
\label{T:ttpopulated}      
\centering          
\begin{tabular}{ccccclc}
\hline\hline
ID  & RA $[\deg]$ & DEC $[\deg]$ & N$\rm_{OB}$ & N$\rm_{TOT}$ & Quad & Notes \\
    & (J2000.0)   & (J2000.0)    &             &             &      &       \\ 
(1) & (2)         & (3)          & (4)         & (5)         & (6)  & (7)   \\
\hline
      151 &   16.260516 &    2.140498 &     99 &    103 & 7,8   &   n   \\
      147 &   16.257084 &    2.158485 &     53 &     58 & 7     &   n   \\
      127 &   16.248152 &    2.153727 &     49 &     54 & 4,7   &   n   \\
      137 &   16.253227 &    2.178541 &     46 &     48 & 7     &   n   \\
      135 &   16.249324 &    2.097876 &     31 &     37 & 8     &   -   \\
      175 &   16.270523 &    2.157452 &     28 &     30 & 7     &   n   \\
      114 &   16.235155 &    2.082594 &     22 &     23 & 5,6   &   n   \\
      103 &   16.229149 &    2.096880 &     19 &     23 & 5     &   n   \\
       21 &   16.135528 &    2.161271 &     17 &     18 & 1     &   n   \\
      185 &   16.276164 &    2.158762 &     17 &     17 & 7     &   n   \\
       56 &   16.203010 &    2.107146 &     16 &     19 & 5     &   -   \\
       44 &   16.186156 &    2.108783 &     16 &     17 & 5     &   -   \\
       88 &   16.217727 &    2.107367 &     16 &     17 & 5     &   -   \\
      122 &   16.242760 &    2.079217 &     15 &     19 &5,6,8,9&   n   \\
      106 &   16.227180 &    2.101775 &     14 &     18 & 5     &   n   \\
      123 &   16.241480 &    2.089006 &     14 &     15 & 5,8   &   -   \\
      132 &   16.248173 &    2.090002 &     13 &     14 & 8     &   -   \\
      191 &   16.283802 &    2.166530 &     13 &     13 & 7     &   n   \\
       76 &   16.213125 &    2.076123 &     12 &     13 & 6     &   -   \\
      154 &   16.256969 &    2.072318 &     11 &     12 & 9     &   n   \\
      161 &   16.262861 &    2.161338 &     11 &     12 & 7     &   n   \\
\hline
\end{tabular}
\end{table}
                      %\label{T:ttpopulated}

%\subsection{General comments}

%Additionally, we found 
%sources with very blue Q but red B-V (usually B-V$>$0).
%These targets are usually located in the central upper locus of the \cc~diagram,
%and do not follow the reddening law. 
%Most of them are galaxies, or extended amorphous sources (probably blends).
%A small fraction of them are isolated red stars with blue
%Q because of the non-monotonic behaviour of the function at spectral types later than B9.

%It is suprising that no OB stars nor
%associations are found in the bar,
%eventhough our criteria for target selection was very
%relaxed in both V and B-V in order to include very reddened stars.
%A possible explanation is that reddening is
%very high in this region, obscuring the stars
%by more than $\sim$5 magnitudes in V.

\subsection{Comments on Individual Associations}
\label{ss:comments}

%\textit{Assoc. \#2}: its most massive star follows the \age 7.4 isochrone.
%
%\textit{Assoc. \#3}: its most massive star  follows the \age 7.1 isochrone.

%\textit{Assocs. \#5 and \#6} are close to each other and would make 
%a single association if \ds~ was slightly longer.
%Association 6's most massive member is on the \age 6.8 isochrone.

%\textit{Assoc. \#8}:
%The brightest object has colors of a VB\_R star, and does not
%follow the reddening law.  
%\mup~ was calculated using the second brightest star, instead.

\textit{Assoc. \#10}: an A3~II (\citet{Fal07}'s C3)
is located 6$\arcsec$.2 away from one association member.

%\textit{Assoc. \#11}: \mup~ is the average derived from the \age 
%7.1 and 7.4 isochrones.
%
%\textit{Assoc. \#13}: its most massive star follows the 7.4 isochrone.

%\textit{Assoc. \#14} displays a circle-like distribution.
%It could be associated with \#15.
%The latter seems slightly older, but the color-magnitude
%diagrams are not conclusive.

\textit{Assoc. \#16}: 
%the color-magnitude diagrams suggest that it formed in 2
%different episodes. 
Part of the association shows
radial distribution around one red and two blue central stars,
that were not included in the OB catalogue because of 
their photometric errors.
%Association \#16 may form a larger group with \#19 (discarded because
%it only has two valid members), \#21 and \#22. 
%Their ages are consistent with this hypothesis.

%\textit{Assoc. \#18}:
%The brightest star is not used for age determination
%because of its very blue Q that puts the star on the \age 6.8 isochrone
%(we suspect that Q color is contaminated).

%\textit{Assoc. \#19}:
%The brightest star, with blue Q, is actually red.
%We suspect that the reason is that Q and spectral type
%are not univocally related at this Q, therefore
%the target is discarded.
%However, its colors are very similar to those
%of the LBV-candidate in association \#158.
%The star would then be \age 6.8 old,
%with 30M$\rm _{\odot}$.   % sin corregir por reddening

\textit{Assoc. \#21}:
%members are distributed in two age groups of
%\age 6.8 and 7.4. The association contains
%one bright red star whose position in the \qv~
%diagram is consistent with an RSG of \age 7.4.
The brightest blue star is a B1~Ia
whose spectrum displays  strong \halpha~ wind emission
(\citet{Fal07}, id: C6).

\textit{Assocs. \#27 and \#28}:
their members are horizontally aligned in our images
(they have very similar RA), perpendicularly
to a spike of a nearby saturating bright star. The brightest stars
of both associations fall right on the spike, and we 
discard them because their photometry may be altered. 
%They are close to association \#25 (discarded).

%\textit{Assoc. \#28}:
%Its brightest member is also located on one spike of the saturating star,
%and we discard it for age determination.
%Close to association \#25 (discarded).
%
%\textit{Assoc. \#31} is also close to the 
%bright saturating star.
%\mup~ was calculated with the \age 7.4 isochrone.
%
%\textit{Assoc. \#34}:
%The two brightest stars are blended and are discarded.
%
%\textit{Assoc. \#35}:
%The brightest OB member
%is actually red, hence discarded.
%
%\textit{Assoc. \#41}:
%The provided \mup~ is the average from \age 7.4 and 7.7.

\textit{Assoc. \#43}: The brightest
star is \citet{Fal07}'s B11, an O9.5~I.
%\mup~ is the average of the values yielded by 
%the isochrones with \age 6.5 and 6.8.

%\textit{Assoc. \#44} is near
%%the \qv~ diagram hints that there are two populations,
%%one with \age 6.5-6.8 and another
%%of 7.4.
%%$[$AFF2006$]$-2047-2 cepheid is included in this association.
%associations \#43 (8$\arcsec$ away) and \#48 (7$\arcsec$ away).

%\textit{Assoc. \#47} is close to association
%\#46 (discarded) and \#50.

\textit{Assoc. \#48}: 
%has three well defined populations
%with \age 6.5, 6.8 and 7.1.
%Visual inspection reveals a blue gaseous filament that could contaminate
%the Q color and shift the location of the targets to the blue in the \qv~ diagram.
%% corresponding to hodge90 hii-13.
%Spitzer observations at 8$\rm \mu m$, which traces
%PAH and hot dust, found a peak in the area covered by
%associations \#43, \#44 and \#48.
Blue stars 7$\arcsec$ away in the SE direction
could connect this association with \#51.

\textit{Assoc. \#53}:
% is close to associations \#39 and \#52
%(both discarded).
%$[$AFM2000$]$-V2384 cepheid and 
\citet{Fal07}'s C16, an A5~Ib,
is included as a red member.

\textit{Assoc. \#54}:
%includes cepheids
%$[$AFM2000$]$-V3062 and $[$AFF2006$]$-2231-2.
A bright blue nearby (8$\arcsec$~ away) star is missed for the association.

%\textit{Assoc. \#56}: \mup~ calculated using the \age 6.8 isochrone.
%
%\textit{Assoc. \#58}: \mup~ calculated using the 
%average values derived using \age 7.1 and 7.4 isochrones.

%\textit{Assoc. \#60}'s brightest member is in the top-ten list of
%brightest stars in the galaxy (see Table~\ref{T:ttbright}). 
%The position in the color-magnitude and color-color diagrams 
%is consistent with that of an LBV. However,
%has a very bright
%member (V$\sim$18) with blue Q color,
%that could be an LBV from its position in the CMD.
%It cannot be discarded
%that it is a foreground star.

\textit{Assoc. \#61}:
Its brightest member is \citet{Fal07}'s B8 (B8~Ib).
%\mup~ is the average derived from the \age 
%7.1 and 7.4 isochrones.

%\textit{Assoc. \#63} contains cepheid $[$AFM2000$]$-AFM2000-V1095.
%
%\textit{Assoc. \#64}:
%\mup~ is the average derived from the \age 
%7.1 and 7.4 isochrones.

%\textit{Assoc. \#68} is close to association \#62 (discarded).
%\mup~ was calculated using the \age 7.1 isochrone.

\textit{Assoc. \#69}: 
%exhibits a spread in age in the \qv~ diagram,
%with possibly three burst at 6.8, 7.1 and 7.4.
The brightest star is \citet{Fal07}'s B10 (B8~Iab).
%The $[$AFM2000$]$-V0314 and  cepheids are also included in the association.

%\textit{Assoc. \#73} is close to associations \#66 and \#69.
%%It includes the $[$AFF2006$]$-8103-2 cepheid.

\textit{Assoc. \#75}
includes an A5~II (\citet{Fal07}'s A2). A nearby red star
(not listed under the association) could be a red supergiant (RSG).
%8$\arcsec$ away from association \#80.

%\textit{Assocs. \#76 and \#77} are only 7$\arcsec$ away. 
%%\#77 has some age dispersion,
%%and its members are resolved in VIMOS images but not in INT 
%%We consequently mark it as a dubious association.
%%\mup~ was calculated from the \age 7.1 isochrone.

%\textit{Assoc. \#79} is close to associations \#73, \#82 and \#86.

%\textit{Assoc. \#80}:
%According to the \qv~diagram,
%the population of 'blue' and 'red' members have different
%ages. 
%The red members look white and bright,
%perhaps they are foreground A-stars. 
%This association is close to \#75.
%\mup~ was calculated from the \age 7.1 isochrone.

%\textit{Assoc. \#81} is surrounded by \#73, \#79 and \#86.
%The $[$AFM2000$]$-V0426 cepheid is included in the association.

%\textit{Assoc. \#82} is close to associations \#79 and \#86.
%It has a nice \cmd~diagram that seems to follow the
%turn-off .  
%The brightest blue stars seem blended in 
%the INT images, but they are resolved in WFPC2.
%\mup~ was calculated 
%with the least reddened star.

\textit{Assoc. \#83}'s brightest blue member has very blue Q.
It is a Be star
(\citet{Fal07}'s B19); its emission lines may explain
its bluer color. Additional faint blue stars are found nearby
(at $\sim$ 7\arcsec-9\arcsec).

%\textit{Assoc. \#84} is surrounded by \#80, \#88, \#92 and \#94.
%%\mup~ was calculated from the \age 7.4 isochrone.

\textit{Assoc. \#85}: An additional blue star is
only 7\arcsec~ away.

\textit{Assoc. \#86}:
% is close to associations
%\#79 and \#82.
Its brightest member  looks triple in the WFPC2 images.
A nearby bright red star could be an RSG star of this association.
%The location of a nearby bright red star
%in the \qv~diagram make it a candidate RSG star of this association.
%$[$AFM2000$]$-V4004 and V1835 cepheids are also in the vicinity.

%\textit{Assoc. \#87} is less than 7\arcsec~ away from association \#85.
%%\mup~ was calculated from the \age 7.1 isochrone.

%\textit{Assoc. \#88} is close to association \#94 and to the bar.
%Its \cmd~diagram is bimodal, one set of
%stars follows the main sequence and the other has redder B-V colors
%suggesting additional extinction.
%However, the redder stars are actually blends.
%%Four $[$AFM2000$]$ cepheids are nearby.
%%For \mup~ determination the second brightest star was used, 
%%because it is younger.

%\textit{Assoc. \#91} is close to association \#93.  
%%$[$AFM2000$]$ cepheid V0530 is included in association.
%Three of the members are close to the \age 6.5 isochrone,
%and two of them on the \age 7.4 isochrone.

%\textit{Assoc. \#93} is close to association \#91,
%and has a nearby cepheid $[$AFM2000$]$-V0580.

%\textit{Assoc. \#94}: 
%A nearby star ($\rm \sim 7\arcsec$ away) could bridge this association to 
%\#92 (which was discarded).

\textit{Assoc. \#96}:
 \citet{Fal07}'s A3 (A7-F0~Ib) is only 7\arcsec away.

\textit{Assoc. \#97}:
%exhibits age dispersion in the \qv~ diagram but not
%in the \cmd~diagram. There is nebulosity
%around some stars, which may explain the dispersion in the former.
\citet{Fal07}'s A4 (B9~II) is only 9\arcsec away.

\textit{Assoc. \#98}:
% is only 6.5\arcsec away from \#95. 
An additional blue target 9\arcsec~ away is missed for the association.
%%The star with the smallest reddening was used for \mup~ determination.

\textit{Assoc. \#99}: 
A member with very blue Q has the same position in the \ubq~diagram
as IC~1613's WO.
%We discard this target for age determination but is is obviously
%follow-up spectroscopy would be very interesting.
However, this object may be blended with the brightest association member.
%but it is unlikely to contaminate the detection because of the difference in magnitude.

\textit{Assoc. \#103}: 
%after cleaning the blends,
%two trends are clearly seen in the \qv~diagram, one with $\log age$ in the 6.8-7.1 interval, and the other 
%with \age 7.4. 
\citet{Fal07}'s A4 star, a B9~II, is included in the association. 
Because of its late spectral type,
this object's Q color is slightly beyond our Q constraint for blue stars (i.e., Q$>$-0.4).
%Association \#103 is close to \#96, \#106 and \#112. 
%The provided \mup~ is the average from \age 6.8 and 7.1.

\textit{Assoc. \#104} 
%is located in the bar.
%, hence differencial
%reddening is likely responsible for 
%the spread seeing in the \cmd~ and \qv~diagrams.
contains \citet{Fal07}'s C17 star, an early-A~II.
The target enters the catalogue as a red star (Q $>$-0.4).

\textit{Assoc. \#105} contains \citet{Fal07}'s A5 star (B3~Ib).
(This association was discarded because it only has 2 valid members).

\textit{Assoc. \#106}:
%close to \#102, 
%on the side of the bar.
%%Most of its blue members have ages in the 7.1-7.4 interval.
%%Some red stars are found in the
%%8.0 isochrones.
The brightest star is \citet{Fal07}'s A4b star (B5~Ib).

%\textit{Assoc. \#108} is close to association \#102.

%\textit{Assoc. \#109} is close to association \#98.

%\textit{Assoc. \#111} is close to the bar,
%and surrounded by a circle formed by associations \#106, \#107 and \#113.

%\textit{Assoc. \#112} is close to \#103 and \#115.

%\textit{Assoc. \#113} is located in the bar, close to \#111.
%Extra-reddening is clearly seen in the \cmd~diagram
%as an offset.

\textit{Assoc. \#114} is a round association located at the outer part of the galaxy.
%After cleaning the \qv~diagram from a
%vertical sequence of stars with redder Q
%which are blends, there is a very well defined
%sequence of stars following the 7.1 isochrone.
The brightest blue star (V$\sim$18.5) could be blended with
the next two brightest stars of the association, although targets
are well resolved in VLT-VIMOS images.
%%We do not use the brightest star to determine \mup~ in case it
%%is a foreground object.
%The location of this star on the color magnitude and color-color
%diagrams is similar to the LBV-candidate in \#158.
%Its age would be \age 6.8-7.1 and would have 
%an initial mass of M=22.9$\rm M_{\odot}$.  % sin corregir por reddening

\textit{Assoc. \#116}:
% is close to \#114, \#115 and \#123. 
A B0~Iab star (\citet{Fal07}'s B14) is close to the association.

\textit{Assoc. \#117}:
The position of the red member in the color-magnitude
diagrams matches that of an RSG.
Close to association \#114, together they could form a ring.

\textit{Assoc. \#120}:
The brightest member is \citet{Fal07}'s A7, B2~Iab. 

%\textit{Assoc. \#121} is close to \#119.

\textit{Assoc. \#122}: 
The brightest blue member is \citet{Fal07}'s B16, B1.5~Iab. 
%The brightest red member may be a red giant of \age 7.7.
%This association is close to \#117.

%\textit{Assoc. \#123}: Flanked closely by associations \#124 and \#126.
%Two well defined trends are seen in the \qv~diagram,
%following  \age 6.8-7.1, and \age 7.4 isochrones.

\textit{Assoc. \#124}'s
brightest member is an A0~III (\citet{Fal07}, id: B15).

\textit{Assoc. \#125} : 
%is located in the bar. %and extra-reddening is seeing in the \cmd~diagram.
Its brightest member could be contaminated by a nearby red star,
although its normal colors suggest otherwise.

\textit{Assoc. \#126}:
% is close to \#123.
The brightest star has very  blue Q 
and its colors deviate from the reddening law,
but  visual inspection
reveals that it is a normal isolated star.
This star is $\rm [AM85]3$, with spectral type
OB+em \citep{ALM88}. The nebular emission could explain
its anomalously blue Q.

\textit{Assoc. \#127}
delineates the bright part of one of the bubbles of the NE lobe.
%The blue members follow nicely the \age 7.1 isochrone in
%both color-magnitude diagrams, with some dispersion
%at the base.
%%The red members follow different isochrones; 
%%some are blue in the RGB composite images, suggesting perhaps
%%extra reddening.
%Some stars show signs of nebular contamination.
\#127 may form a single giant association with \#147.
%Other nearby associations are \#128, \#130 and \#136.
Spectral types are known for three stars in the association:
the brightest blue stars have types
A2~Ia and B1.5~Ia (\citet{Fal07}'s id's
are A8 and B4). 
%Note that the nearby association 128 contains a Be star
%(\citet{Fal07}'s B5).
\citet{S07} assigns A0~Ia(e) to the former
and additionally lists an RSG, $\rm [S71]$V32 with type M1~Ia 
(not included in our catalogue because of the hard constraints on blue photometric errors).
The A-supergiant %[S71]A43
has been suggested being a foreground
star \citep{SK76}, but \citet{Fal07} found its radial velocity in agreement
with IC~1613, and report P~Cygni profiles of 
$\rm H_{\gamma}$ and $\rm H_{\beta}$ (see Sect.~\ref{sss:Qchart}).
%Higher resolution spectroscopy is needed to determine 
Spectroscopy of higher resolution than \citet{Fal07}'s VLT-FORS2 data
is needed to determine
whether this object is an LBV-candidate.
%\mup~ listed in 
%Table~\ref{T:ASSprop} corresponds to the second brightest star, the B1.5Ia,
%located on the \age 6.8 isochrone.

\textit{Assoc. \#128}:
% is very close to association \#127. 
The brightest star is a Be (\citet{Fal07}'s B5), and this
may explain its very blue Q color.
%Its located on the \age 6.5 isochrone, and its initial
%mass would be 32$\rm M_{\odot}$.  % sin corregir por reddening
%%Since Q is not reliable, we
%%do not use this star for \mup~ determination.
%%Nebulosity is seen on the second brightest star.
%%Listed \mup~ is the average of the masses derived using
%%\age 7.1 and 7.4 isocrhones.

%\textit{Assoc. \#130} is close to \#127.
%%A red member is located on the 8.0 isochrone; it probably
%%belongs to the underlying old population.

%\textit{Assoc. \#131} is surrounded by
%\#120, \#126, \#129 and \#135.

%\textit{Assoc. \#132}:
% is a well populated association.
%All stars follow the \age 7.4 isochrone in the \cmd~diagram,
%but in the \qv~diagram the members are separated onto
%the \age 6.5 and 7.1 isochrones.
%There are two candidate blue supergiants
%with ages 6.5 and 7.1 respectively.
%%For \mup~ determination we use the second brightest
%%star, which follows the youngest isocrhone.
%One of the red members is on the locus of RSGs.
%%There is a cepheid in the association,
%%$[$AFF2006$]$-1450-2.

%\textit{Assoc. \#133}: its most massive star  follows the 7.1 isochrone.

\textit{Assoc. \#134}:
A possible additional member, bright and blue, is
only 7\arcsec away.

%\textit{Assoc. \#135} is close to \#131 and to the bar.
%%The brightest star has intermediate \age 6.8-7.1,
%%and \mup~ was calculated consequently.

\textit{Assoc. \#136} 
%is  close to association \#127 and to the bar.
marks a bubble rim.
%The brightest star has intermediate \age 7.1-7.4,
%and \mup~ was calculated consequently.

\textit{Assoc. \#137} 
%The \qvd~ hints 3 episodes of star formation at \age 6.5, 7.1 
%(including an RSG candidate) and 7.4.
is formed by several
arc-like structures that possibly trace the edges of several
bubbles.
Four members at the West could
form a separate association. %of \age 7.7.
The brightest member is a B supergiant:
B1~Ia (\citet{Fal07}'s A10), B2/3~Ie \citep{S07},
B2-3~I \citep{Hu80}.
The second brightest star is 
\citet{Fal07}'s A9 (B5~Iab).

\textit{Assoc. \#140}: 
%A nearby cepheid $[$AFF2006$]$-3384-2 and an
A nearby blue star ($\sim$7\arcsec~ away) is not included in the association.
%The most massive star is on the \age 7.1 isochrone.

%\textit{Assoc. \#143}: its most massive star follows the \age 7.4 isochrone.

\textit{Assoc. \#144}: 
%%age cannot be determined because
%%2 out of 3 members have very blue Q color
%%and do not follow the reddening law.
%%%, suggesting contamination by nebular emission.
%%The third one has \age 7.1.
%The association is located in the gas rich bubble
%region. 
The brightest member
ionizes its surrounding nebula.

%\textit{Assoc. \#145} is close to association \#132, in the
%outer part of the galaxy. 
%%One faint member has red Q, but visual inspection does not show evidence
%%of a blend or nebular contamination.
%%The remaining explanation,
%%extra reddening, is unlikely
%%because of the location of the association in the galaxy.
%%%\mup is the average value produced using isochrones 
%%%with \age 6.8 and 7.1.

\textit{Assoc. \#146}:
% is close to the bar and to \#155 and \#160.
%%Some members follow
%%the \age6.5 isochrone, and the brightest one is a 
The brightest member is a Be (\citet{Fal07}, id: B12). 
%%There is an additional older
%%sequence, following the \age 7.4 isochrone,
%%whose brightest star may be evolving towards the red
%%part of the color-magnitude diagram.
%%%\mup~ was calculated with the Be star.
%%%(that may have wrong Q, hence wrong Av correction).

\textit{Assoc. \#147} 
%is close to associations \#157 and \#161.
is dominated by 2 clumps, each 
in the intersection of 2 bubbles, plus 2 arc-like structures
that trace the rim of one bubble each.
%Since there is overlap with ionized gas, 
%Q is likely bluened by nebular lines.
Some detections are actually nebular knots (discarded).
The association has 5 bright members ($\rm V \in [19,20]$),
four of them in the largest clump of stars.
\citet{Lal02} obtained low resolution spectra ($\rm \Delta\lambda=$8\AA) of the
stars in the main clump, with a 6m telescope (note
that this area is hardly resolved, even  in the VLT-VIMOS images).
The group includes an Of star, two O~giants
and two OB candidates.
The association has an additional OB member,
far from the regions of crowding (OB+em?, \citet{ALM88}).
%%For age determination we discarded two of the brightest stars, 
%%with redder B-V, which are actually blends.
%All 5 isochrones from \age 6.5 to 7.7 are well populated.

\textit{Assoc. \#148}:
% is located in the outer part of the galaxy, 
%between \#145 and \#156. 
A nearby blue star, potential member of the association,
is 7.5\arcsec away.
%\mup~ calcuated with the second brightest star,
%because it is younger.

\textit{Assoc. \#149} is in the bubble region, between the large associations
\#137 and \#147, but no nebulosity is seen nearby.
%Close to associations \#143 and \#158.

%\textit{Assoc. \#150} is a trapezium-like association,
%close to the bar and to \#151.

\textit{Assoc. \#151} %is young and well populated.
%Age spreads from
%\age 6.5 to 7.7, with higher concentration in the interval \age 6.8-7.1.
%%The red members are an unrelated population of \age 8.0.
is the association with more OB members of our catalogue.
%is located in the bubble region,
%closely surrounded by \#150, \#153, \#157 and \#171.
It is arc-shaped and part of it follows one bubble.
IC~1613's known supernova remnant (SNR) is included in the association.
Our catalogue yields 5 detections for the SNR, which we discard. 
%that occupy the locus of  VB\_B stars.
We note here one apparently isolated target 
%in the locus of WR stars (Q$\sim$-1.4).
whose position in the Q \textit{vs} U-B and B-V \textit{vs} U-B
color-color diagrams is the same as for the WO star.
%\citet{AM85}, in a search of WR from their emission lines, found
%an object close to the SNR, WR-8. 
%\textbf{comprobar si es la misma deteccion!: NO, no es la misma}
%\citet{MCA87} reported 
%HeII4686 emission, broader than nebular but still weaker than expected
%for a WR. However, from improved data \citet{AM91} found also iron lines
%and showed that 
%the HeII4686 emission lines are not broader than nebular,
%and that they actually resembled emission from the Crab Nebula 
%and Cygnus Loop filaments, indicating that this is a young SNR.
Several obscured clouds fall within the limits of the association.

%The SNR could provide constraints on the age of the association,
The age of the SNR is yet undecided since it
 shows both young and old signatures.
While it was originally believed to be about 2E4 years old  \citep{DD83,PBT88}
X-ray observations suggest an age of 3E3 years \citep{Jal98}.
The emission lines from the knots resemble those from the Crab Nebula
and Cygnus Loop filaments, indicating that this is a young SNR \citep{AM91}.
It has been suggested that the SN exploded inside a large cavity surrounded
by a dense \ion{H}{i} cloud. The remnant would have just encountered the dense material,
producing shocks and hence the peculiar features \citep{Lal98,LMP03}.

Spectral types are known for three stars in this association.
The two brightest blue stars are an O9~I  and a B1~Ia, \citet{Fal07}'s B7 and B6.
The authors report strong \halpha~ wind emission in the spectrum of B6.
The association contains a red supergiant,
%There is a red supergiant with \age 7.1 in the association,
\citet{SK76}'s variable star $\rm [S71]$V38 (M0~Ia, \citet{S07}).
The RSG is only 6.7\arcsec~ away from association \#150.

%The most massive star is the third brightest star,
%and lies on the \age 6.5 isochrone.

%\textit{Assoc. \#153}:
%the brightest star, used for \mup~ determination, may be a blend.

\textit{Assoc. \#154} encloses IC~1613's known WO star.
%The target is in the VB\_B category and its colors
The object has very blue Q and U-B colors
%, and does not follow the extinction law 
(V=19.86, Q=-1.35, B-V=0.02, U-B=-1.33).
The WO is located on the edge 
of the giant \ion{H}{i} shell that contains the SW side of the galaxy,
and ionizes an asymmetric bipolar \ion{H}{ii} region \citep{Aal00,Loal01}.
These authors find % with narrow-band imaging in heII,halpha,[SII]6717+6731,[OIII]5007
at least two extra sources in $\rm H_{\alpha}$ and HeII~4686 narrow-band filters
that are compact knots of gas.
% because the smallest  lobe hits the dense \ion{H}{i}, while the large one expands freely.
Some entries of our OB candidate list are also
knots of the nebula ionized by the WR.
They are characterized by very blue Q and red B-V.
However, U, B and R VLT-VIMOS images
suggest that two of the detections (besides the WO) are in fact stars. 
One of them has very blue Q, and the other one, very embedded in
the WO nebula, has red Q.
These stars might be members of a cluster hidden
behind the WO, proposed by \citet{Loal01}.
%, since the continuum of the spectrum is spatially extended.
%Since the Q colors are contaminated by the nebula,
%we cannot determine age.
The bipolar shell's
dynamical age is 0.3-1E6 years \citep{Aal00},
consistent with the duration of the WR stage.

%\textit{Assoc. \#155} is close to \#156 (discarded), at the SE end of the bar.
%In the \qvd, stars separate into 2 groups
%following the \age 6.8 and 7.4 isochrones.
%The group of older stars form an arc in the images,
%and perhaps constitute an older background population.

\textit{Assoc. \#157}:
% is close to association \#147.
VLT-VIMOS images reveal that the brightest star is triple, and
unresolved in our catalogue.
%The star with very red B-V, but blue Q looks actually red. 
One member has similar colors to the brightest star
in \#158, which
%The association includes a B\_R star (blue Q but very red B-V).
%A brighter target in association \#158, with similar colors,
has been found to be an LBV-candidate.

\textit{Assoc. \#158} is located in the void of the bubble region, like \#149.
One of the members has very red B-V and blue Q.
%falling in the B\_R  box of the \cc~diagram.
Medium resolution spectroscopy soon to be published 
(Herrero et al. 2009, in prep.) has revealed P~Cygni profiles of the Balmer series
and of iron, rendering the
target an LBV-candidate.
%The LBV would have an initial mass of $\rm M_{\odot}$.  % sin corregir por reddening
%                        One corrected from reddening (habria q preguntarse si esto es correcto,
%                          pq los colores no tinene pq parecerse a una estrella OB),
%                          M_ini (6.5)=73, (6.8)=30.

%\textit{Assoc. \#160}: 
%Most massive star is on the \age 7.1 isochrone.

\textit{Assoc. \#161} 
%is located in the bubble region of the galaxy.
traces two bubble rims, sharing one of them with association \#147.
%One red member follows the \age 8.0 isochrone.
The brightest target is \citet{Fal07}'s A11 (B9~Ia).
%the images reveal that it could be blended to a red star.
%Listed \mup~ is the average of the masses derived using
%\age 6.8 and 7.1 isocrhones.

\textit{Assoc. \#162}'s
brightest star is in the
intersection with one of the bubble rims followed by \#161.
Its spectral type is O5-O6~V (\citet{Fal07}'s B2),
and it is likely ionizing its surrounding nebula.

%\textit{Assocs. \#167 and nearby \#168} sit in the outer reaches of IC~1613
%at the SE end of the bar.

%\textit{Assoc. \#170}: 
%\mup~ was calculated using the 7.4 isochrone.

\textit{Assoc. \#175}:
%  is close to \#173, \#176 and \#183.
%It exhibits a well populated blue plume in the
%\cmdd, with ages ranging from \age 6.8 to 7.4.
%The red population follows the \age 7.7 and 8.0 isochrones.
The association is spatially divided into two parts,
one at the centre of a bubble,
%one blowing material away and likely creating
%the enclosing bubble.
%The other part is in the intersection of the resulting
%bubble, and another one blown by association \#185.
and the other in the intersection of such bubble and 
another that encloses association \#185.
%All members will experience nebular contamination,
%hence blue Q.
One apparently isolated member has very blue Q ($\sim$-1.4),
and its position in the
Q \textit{vs} U-B and B-V \textit{vs} U-B diagrams
is similar to the WO.

\textit{Assoc. \#176}: Two out of its three members
are in the centre of a small bubble.
%and its probably them which are blowing the material away.

%\textit{Assoc. \#178}
%includes a bright red star whose position in the color-magnitude
%diagrams agrees with that of an RSG
%between \age 7.1 and 7.4 isochrones.
%%\mup was derived from the star that follows the \age 7.4 isochrone.

\textit{Assoc. \#179}'s brightest star is \citet{Fal07}'s A12, a B1.5~Iab.

%\textit{Assoc. \#180} is in the bubble region
%and close to associations \#179, \#183 and \#188.
%%A nebular veil over the members is seen in the INT-WFC images.

\textit{Assoc. \#184} is found in the SW outskirts of the galaxy.
The stars are distributed in a ring shape, around a central
red star. % of \age 8.0.
%Listed \mup~ is the average of the masses derived using
%\age 7.4 and 7.7 isocrhones.

\textit{Assoc. \#185}: % is a very young region.
Members are located on the rim of
one bubble, but they may also be creating the bubble where part of association
\#175 is located at the SW.
%This is a very crowded region, and even the brightest members could be blends.
The association has a high stellar density and all members may experience blends.
The brightest star is an early B-supergiant
(B0~Ia \citet{Fal07}'s B3; B:I: \citet{SK76}'s B42)
although \citet{Lal02} classified it as an O supergiant.
The study of the latter in this region (see comment on association \#147 
for a description of their analysis)
found 3 more OB stars: O~III, B8~II and one with uncertain type
O8~III or O4~V . %(listed in Table~\ref{T:SpType}).
A faint member (V$\sim$22) with extremely blue Q
and U-B, sits in the same position in the \ubq~ diagram as
the WO.

\textit{Assoc. \#187}'s brightest member is \citet{Fal07}'s A13
a very early O dwarf (O3-4~V((f))),
which is remarkable for a modest association of only 3 members.

\textit{Assoc. \#189} is located on the rim of a bubble.
The spatial distribution suggests that the 
\ion{H}{ii} shell could have been blown either by association \#185 
or by a nearby isolated star.
%The second brightest member is used for \mup~ determination
%since it is younger.

\textit{Assoc. \#191} is linear in shape and
located in the bubble region,
close to  association \#189.
%Nebulosity is seen in the area.
The brightest star is a late-O giant
(O9~III, \citet{Fal07}'s A15).

%\textit{Assoc. \#192}: 
%No nebulosity is seen even though the association is
%near the bubbles.
%There is risk of contamination of the brightest star with a red star;
%the pair is resolved in VLT-VIMOS R-band images.

%\textit{Assoc. \#194} is located East of \#191.
%%The brightest object is discarded, because it is an edge-on
%%background spiral galaxy. 
%%\mup~ corresponds to the second brightest star,
%%with intermediate age between 7.4 and 7.7 isochrones.

\textit{Assoc. \#196}'s brightest star is a Be, \citet{Fal07}'s A17.
Their target A16 (early-B~Ib) is only 6.1\arcsec~ away.
%, and its location 
%in the \qvd~ is consistent with the age assignated to the association.

%\textit{Assoc. \#197}: 
%Despite its blue Q, the brightest target has strange U,V,R colors.
%We discad it for \mup~ determination and use the second
%brightest star instead.
%%This target may be a blend.

\section{THE EFFECT OF THE V-FAINT LIMIT ON THE FINAL CATALOGUE OF ASSOCIATIONS}
\label{ss:qcfa}

The selection criteria for OB stars were flexible in magnitude and colors
to admit very extincted stars,
since we suspected that internal reddening was
significant.
However, after inspecting the associations we have found
that the redder faint stars are not reddened blue stars,
but mostly non-stellar objects or blends.
We consequently wonder how allowing such a large range of V-magnitudes
and colors has distorted the resulting catalogue of associations.
Fainter stars are more abundant, hence closer to each other, and
may shift the value of \ds~ to shorter distances.
By setting \ds~ to a short value, we may be missing %for the association headcount
looser groups of bright OB candidates.

\begin{figure}[]
   \hspace{-1.5cm}
   \includegraphics[angle=90,width=0.7\textwidth]{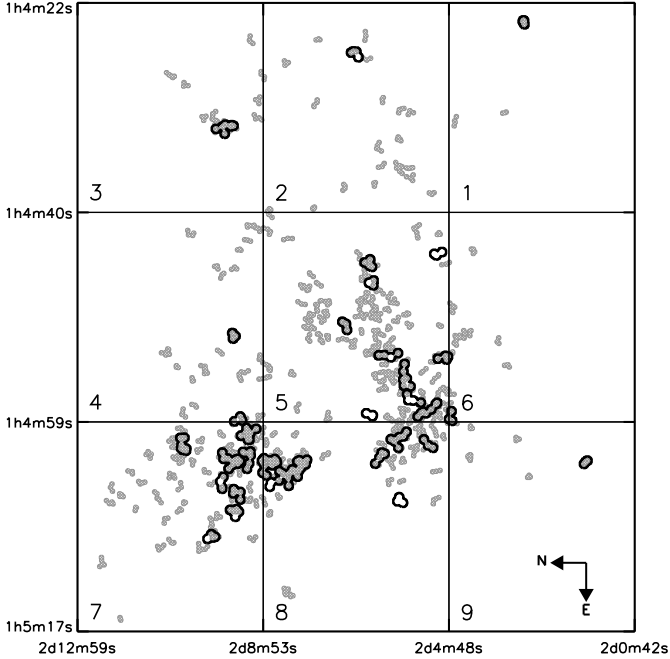}
      \caption{Comparison of the results presented in this paper (catalogue-A, grey shadows)
       and the results of the OB association search algorithm
       run on the same sample of stars but  V$<$21 (catalogue-B, black).
       The associations found in both
       cases agree on the bubble region, but 
       %\textbf{
       the SW side of the galaxy %} 
       is dominated by associations with fainter stars.
       Catalogue-A misses 3 associations
       of bright stars.
       }
       \label{F:comp}
   \end{figure}

We ran the OB association search programme on a
subset of the initial sample: 
bright stars with V$<$21. 
The resulting list of associations (hereafter catalogue-B)
is compared with our results (catalogue-A)
in Fig.~\ref{F:comp}.

As expected we obtained a longer 
\ds=11$\arcsec$ for the less dense group of bright stars.
%whereas for the faint stars we get \ds=7$\arcsec$,
%similar to the value used in this work.
%The associations found when considering only faint V$>$21 stars
%mostly agree with the results of this paper.
%The overlap is not complete, since we have removed bright
%stars from sample A.
There is good agreement of both catalogues
on the bubble NE region of the galaxy.
The associations in catalogue-B
are usually larger and enclose several associations of catalogue-A.
However, 
%\textbf{
the SW side of the galaxy in the \ion{H}{i} cavity %}
is apparently dominated by fainter stars,
and by limiting ourselves to the brightest ones we would have missed
a large number of associations.
On the other hand, by enlarging \ds~ to 11\arcsec~ we would include three more associations.
%but some currently separated associations would merge.
Whether they are % three new associations from catalogue-B are
real OB associations can only be decided after closer inspection
of their stellar content. 

We conclude that the criteria used to build catalogue-A
favoured the identification of OB associations,
although some extra work was needed afterwards to discard
fake identifications.

%These results hint that instead of using a global value,
%\ds~ should be customized
%according to the galactic density at different positions.
%A possible interpretation is that different types of populations
%(i.e. associations \textit{vs} clusters) have different scale lengths.
%\citet{Bal98} have already suggested running the algorithm locally
%to separate different hierarchical groups.
%We further elaborate on this topic
%in the second paper.

%example of missing association:
%68074	68468	68599	form a triangle. they are bright and blue, but further than 6'' from each other.
%the latter is MAUP_B13 (B5Iab)

%\citet{EIN87} proposed a paradigm 
%built from observation of 
%the Local Group (including the MW, M31, M33, Sextans~A and IC1613),
%where stellar groupings with young objects
%($\rm 10^6-10^8 yr$, OB associations, \ion{H}{ii} regions, \ion{H}{i}, WR stars, cepheids,
%RSG and young stellar clusters) can be embedded into 
%succesive larger groups: 
%$\sim$0.2Kpc (aggregate), $\sim$0.6Kpc (complex) and $\sim$1.2Kpc (supercomplex).

\section{COMPARISON WITH PREVIOUS CATALOGUES}
\label{ss:comp}

Four catalogues of OB associations in IC~1613
have been published to date.
The first one
was made from visual inspection of plates and published by \citet{H78}. 
More recent works by 
\citet{I96} (catalogue not public),
\citet{Gal99} and \citet{Bal04}
have used an automatic procedure very similar to ours.
%(see discussion in \ref{ss:parms}).
%They are compiled in Table~\ref{T:comp}.
Our work has produced the most numerous catalogue:
a total of 163 stellar groups,
almost triplicating the 60 associations found by
Borissova et al.
\citet{H78} listed 20 associations and \citet{I96} only found 6.
\citet{Gal99} studied a smaller field.

Direct comparison of our results with the 
seminal work of \citet{H78} can only be done qualitatively,
since his catalogue is marked on low resolution photographic plates of the galaxy.
The associations we have found follow the same distribution
as \citet{H78}'s, but our code breaks them into many smaller ones.
\citet{Gal99} observed the same effect in their results for the
bubble area of the galaxy, and pointed out that their associations
are the bright cores of Hodge's.
We also find new associations both at the core of IC~1613 and
the periphery.
Both facts are explained
by the higher sensitivity and finer spatial resolution of this work.

\begin{figure}[]
   \hspace{-1.5cm}
   \includegraphics[angle=90,width=0.7\textwidth]{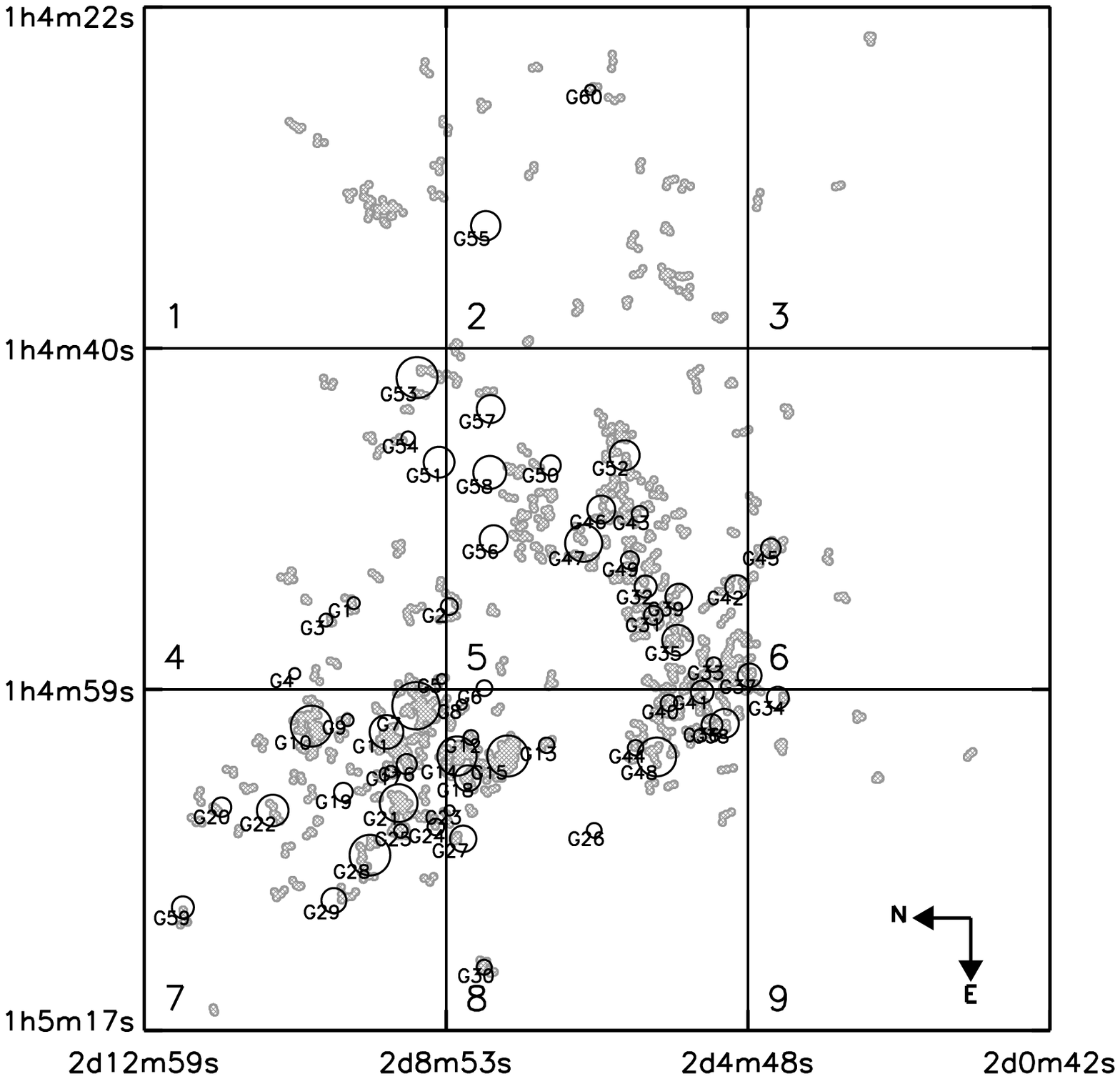}
      \caption{Comparison of the associations found in this work (grey),
       with those of \citet{Bal04} (black).
       The quadrants mark the same positions as in Fig.~\ref{F:all}.
       The overall distribution of the associations in both catalogues agrees.
%       \textbf{
       Considering the associations in the \ion{H}{i} cavity, our results separate \citet{Bal04} groups
       at the SE into many pieces. 
       %However, in the bubble side of the 
       %galaxy is the other way around.
       At the NW end, %
       \citet{Bal04} find 4 associations (G51,G56,G57,G58)
       that do not have a counterpart in our catalogue.
       Other unmatched associations are their G26 and G55.
       }
       \label{F:GvsB}
   \end{figure}

The associations found by \citet{Gal99} are provided on
detector coordinates or plotted over a poor quality image,
and since the same team studied a wider field in \citet{Bal04}
we preferred the latter for comparison.
%A list of the coordinates for the centre and an estimation of the radius
%is provided in Table~1 of their paper, and we compare them with
%ours in Fig.~\ref{F:GvsB}.
We use the coordinates and radius of the associations in their 
Table~1 to compare with ours (see Fig.~\ref{F:GvsB}).
We noted a discrepancy between the centres provided in their list
and the position of some associations they marked over a picture
of IC~1613 in their Fig.~1.
By providing electronically the list of OB members 
of each association, we expect to minimize
these kind of inconsistencies and to make direct inter-comparisons easier.

%It was not possible a direct comparison of our results with 
%previous catalogues of OB associations,
%which are often marked over low resolution photographic plates of the galaxy 
%like the seminal work of \citet{H78}.
%More recent results by \citet{Bal04}, who applied a very similar automatic
%algorithm to find OB associations is also hard because it only provides
%coordinates for the centre and an estimation of the radius.
%One of the advantages of this work  is that 
%we will provide a list of the OB members of each association,
%and also the less massive population in the appendix
%(an example is provided in Table~\ref{T:OBass}),
%making direct inter-comparisons easier.

The overall distribution of our catalogue and \citet{Bal04}'s associations agrees.
The data they used have a plate scale similar to
ours but again, our catalogue separates most of their associations
into two or more, 
%\textbf{
especially in the SW part of the 
galaxy enclosed in the \ion{H}{i} cavity 
%}
(see Fig.~\ref{F:GvsB}).
Our catalogue has additional associations in the outer reaches
of IC~1613, as expected because of our extended field of view,
but also in the overlapping areas.
%Comparison with \ion{H}{i} data (see Fig.~\ref{F:\ion{H}{i}}) has shown that
%it may be expected to find associations in this region.
The likely reason why these associations were not registered by
Borissova et al. is that 
they do not reach the minimum number of elements 
considered in their work (\nm=4).
In other cases the members are faint,
beyond Borissova et al.'s limiting V-magnitude.
%This is probably the reason why we find more
%associations, because we have more targets at the starting point.
These authors found 3 associations (G26, G51, G55) far from any of 
ours (see Fig.~\ref{F:GvsB});
we do find OB stars on those locations, but they are separated
more than \ds~ %(but less than \citet{Bal04}'s \ds)
and hence not included in our catalogue.
There is also a mismatch between associations G57-G58
and the centre of our counterparts.
%the explanation is that they used a slightly larger \ds~parameter.

The different number of associations found by
\citet{H78}, \citet{Bal04} and ourselves
may be better explained in light of the experiment described in
Sect.~\ref{ss:qcfa}.
Hodge only considered stars with V$<$21 to find associations,
and Borissova et al. have completeness problems for fainter magnitudes.
% catalogue-A. V>21, 182 assoc, average radius 6.2'', ds=7''
% catalogue-B. V<21, 31 assoc, average radius 8.8'', ds=11''
Catalogue-B consists of 31 associations of average
diameter similar to the mean size of the associations of \citet{Bal04},
decreasing the discrepancies.
However, catalogue-B still does not include
\citet{Bal04}'s associations without counterpart in catalogue-A.

\section{SUMMARY AND FUTURE WORK}
\label{s:conclusions}

This paper presents a deep multi-band photometric catalogue
of IC~1613 with high astrometric accuracy.
Using the univocal relation between the reddening-free
Q parameter and OB spectral types, we have built a 
list of candidate young massive stars.
With a friends-of-friends algorithm developed by
our group, we have studied how they cluster into associations.
A total of 163 associations were found 
(after discarding 5 galaxies and 29 cases with less than \nm~ valid members), 
ranging from
small groups of 3 stars to large complexes of more than 20 members in the 
centre of \ion{H}{ii} shells.
24 associations have 10 or more OB candidates.
%\textbf{
Our findings are dependent on the choice of the
two free parameters of the friends-of-friends algorithm:
the search distance (6\arcsec) and the minimum number
of OB stars defining an association (3).
%}

The resulting set of candidate OB stars is
ready to be observed with multi-object spectrographs.
To achieve the required astrometric precision,
a new solution for the WFC at the INT was developed.
%This work also provides some insight into
%the enclosing population, which will lead
%to the choice of more interesting stars for 
%the observations.

The photometric analysis leaves an interesting
by-product, an extinction map of IC~1613. Most of
the OB stars exhibit a color excess larger than
the foreground value. Extinction is not uniform
and does not have a smooth distribution. The points of 
enhanced reddening are spread throughout the
galaxy and show a slightly higher concentration at the 
%\textbf{
edge of the \ion{H}{i} cavity that contains the SW side of the galaxy.
%}

The positive confrontation of our list of candidate OB stars
with published spectral types, endorse our target selection
criteria based mainly on Q and secondarily on U-B colors.
Four stars are found in the same position of the \ubq~diagram
as the known Wolf-Rayet star of the galaxy.

This article is the first effort towards a 
comprehensive description of the massive stellar population
of IC~1613, as a proxy for 
the low-metallicity early-Universe.
The next step is the characterization of the physical
properties of the associations presented here,
especially mass and age.
The study of the color-magnitude and color-color 
diagrams of the catalogue associations
and their comparison with theoretical
isochrones will be presented in  part-II.
It will provide interesting clues 
about star formation processes of massive stars
and their IMF in low metallicity environments.

\begin{acknowledgements}
%\textbf{
The anonymous referee is thankfully acknowledged
for his/her useful questions and comments. %}
The authors also recognize
P. Stetson for providing the
photometry of IC~1613.
S. Sim\'on-D\'{\i}az is thanked for his careful
reading of this manuscript.
This work has been partially funded by Spanish MEC
under Consolider-Ingenio 2010, programme grant
CSD2006-00070 (http://www.iac.es/consolider-ingenio-gtc/).
The authors gratefully acknowledge support from
the Spanish Ministry
of Science and Innovation, grant numbers
AYA2007-67456-C02-01 and AYA2008-06166-C03-01.
\end{acknowledgements}

%%%%%%%%%%%%%%%% BIBLIOGRAPHY %%%%%%%%%%%%%%%

%%%%%%%%%%%%%%%%%%%%%%%%%%%%%%%%%%%%%%%%%%%%%

\Online 
\begin{appendix}

%%%%%%%%%%%%%%%%%%%%%%% MEGA TABLA PROPIEDADES ASSOC %%%%%%%%%%%%%%%%%%%%%%%%%%%%%%%%

\section{CATALOGUE OF OB ASSOCIATIONS}

\subsection{Complete Catalogue}

    The output list of OB associations provided by our code
is presented in Table~\ref{T:ASSprop}.

The content of different columns is:
\textbf{(1)} OB association identification number. \textbf{(2)} and \textbf{(3)} Right Ascension (RA)
and Declination (DEC) of the centre
determined from the average of the positions of the OB members, equinox J2000.0.
\textbf{(4)} Number of OB members ($\rm N_{OB}$) of the association.
\textbf{(5)} Total number of members ($\rm N_{TOT}$);
this figure may not be representative since
our photometric selection criteria is biased against red objects
by demanding high U-magnitude accuracy.
\textbf{(6)} Quadrant allocation in  Fig.~\ref{F:all}.
\textbf{(7)} Notes: 'Gal' -- the association is actually a galaxy; 'd' -- 
association discarded because it does not have enough valid members
(i.e. visual inspection reveals
that some are blends or gas knots);
'?' -- dubious association (associations with a small number of members,
with a significant fraction under suspicion of being blends or not blue
stars);
%'?' -- dubious association (a significant fraction
%of members are under suspicion of being blends or not blue
%stars);
'n' -- an entry 
in Sect.~\ref{ss:comments}.

%{\tiny
%\addtocounter{table}{5}       % table moved to after RFs
%\input{TABLES/T_OB}           % label {T:OBass}
%\addtocounter{table}{6}       % table moved to after RFs
%\input{TABLES/T_ASSOC_sum}   % label {T:ASSprop}
%}

\subsection{Member Listing Example: Association \#~16}

Our code provides us with a list of the
OB candidate members of each association and also the red population.
Defining OB associations by their members, rather than by hand-drawn
lines on photographic plates, allows a direct comparison of results between different
works and with observations.
% (if, for instance, we want
%to compare the distribution of OB associations with \ion{H}{i} maps).
The member listings of all the associations of our catalogue
will be available on-line.

As an example, we list the members of association \#16 in Table~\ref{T:OBass}.
The content of different columns is:
\textbf{(1)} Star identification number in our catalogue. 
\textbf{(2)} and \textbf{(3)} Right Ascension
and Declination, equinox J2000.0.
\textbf{(4), (6), (8), (10)} and  \textbf{(12)} magnitudes.
\textbf{(5), (7), (9), (11)} and  \textbf{(13)} error of magnitude measurements.
\textbf{(14)} and \textbf{(15)} DAOPHOT's $\chi$ and \# parameters.
%\textbf{(16)} Chip of the INT-WFC where the star was detected.
\textbf{(16)} Identification number in the 2MASS catalogue.

\subsection{Finding Charts of the Associations}
\label{S:f-charts}

\begin{figure*}[!h]
%   \centering
%   \hspace{-1.75cm}
   \includegraphics[angle=180]{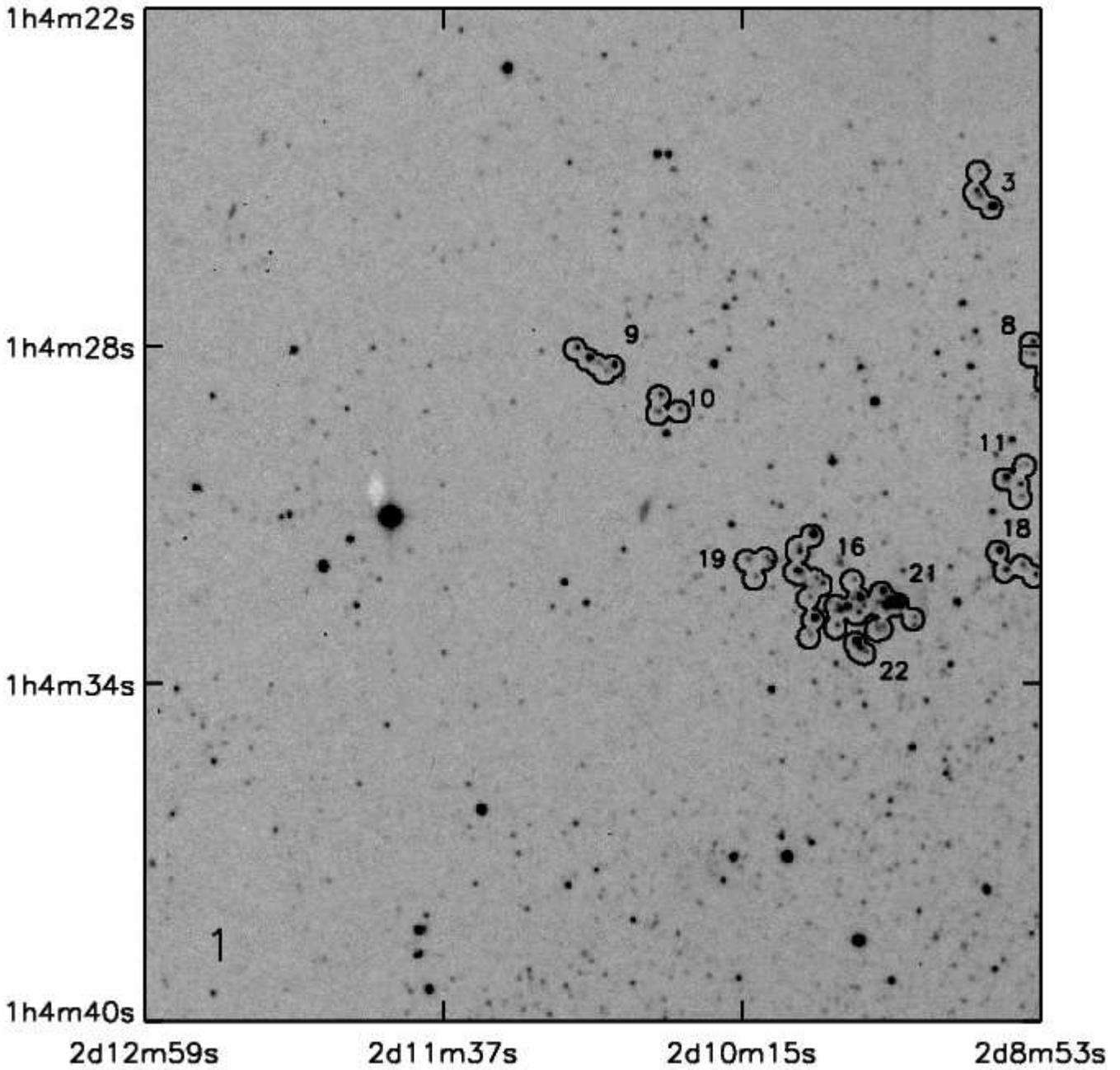}
      \caption{INT-WFC U-band image of a portion of IC~1613 corresponding
to quadrant~1 of Fig.~\ref{F:all}. 
The OB associations located in this quadrant are
indicated by a black contour. Their numbers are:
\#3, \#8-\#11, \#16, \#18, \#19, \#21 and \#22.
%%%%%%% COLOR F_CHARTS
%The image is an RGB composition
%of  U- (blue), V- (green) and R-band (red).
%The OB associations located in this quadrant are
%indicated by a red contour.
}
       \label{F:f-charts1}
   \end{figure*}

\begin{figure*}[!h]
%   \centering
%   \hspace{-1.75cm}
   \includegraphics[angle=180]{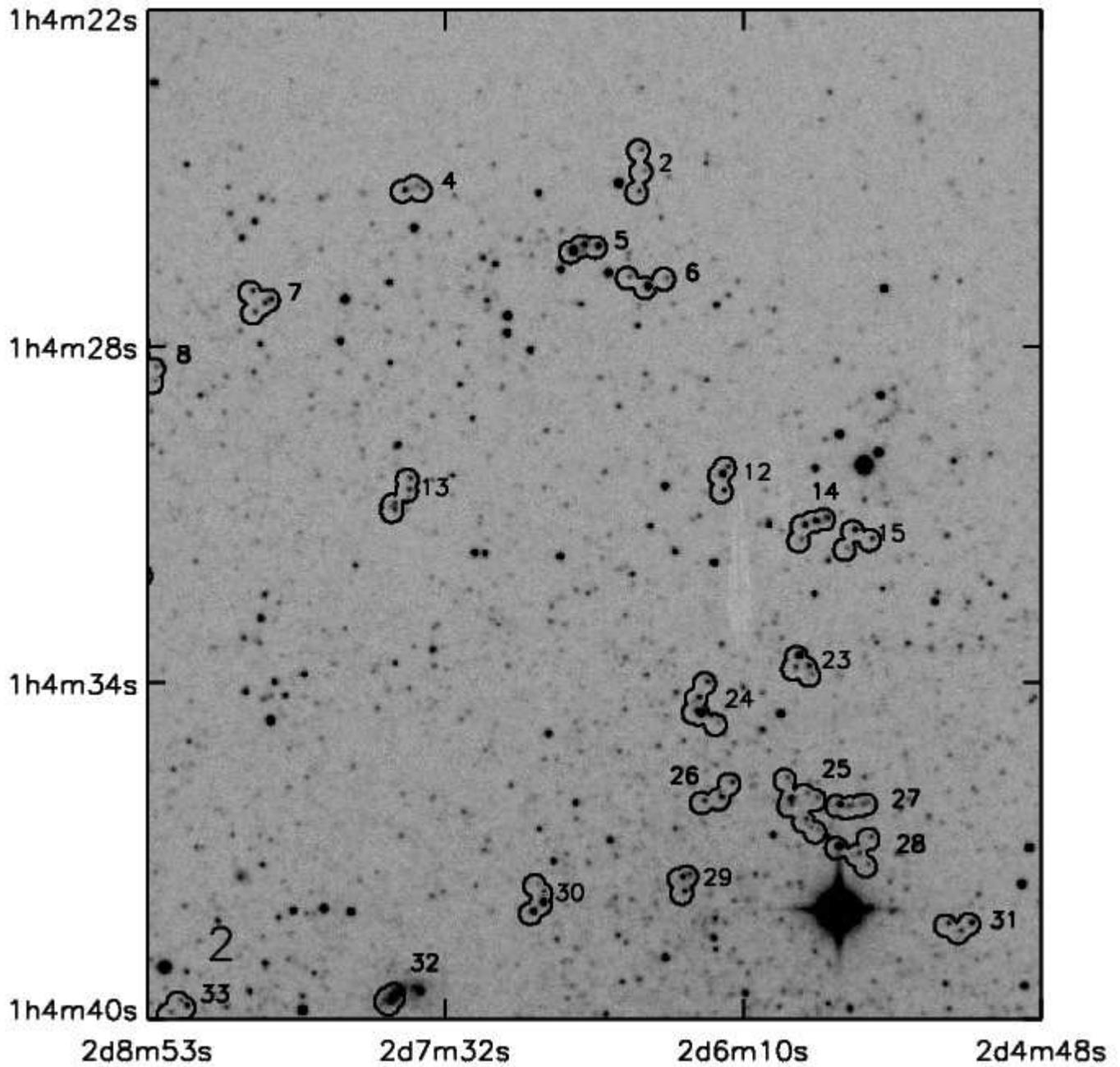}
      \caption{Same as \ref{F:f-charts1}, quadrant~2.
Associations present in this quadrant:
\#2, \#4-\#8, \#12-\#15, \#18, \#23-\#33.
}
       \label{F:f-charts2}
   \end{figure*}

\begin{figure*}[!h]
%   \centering
%   \hspace{-1.75cm}
   \includegraphics[angle=180]{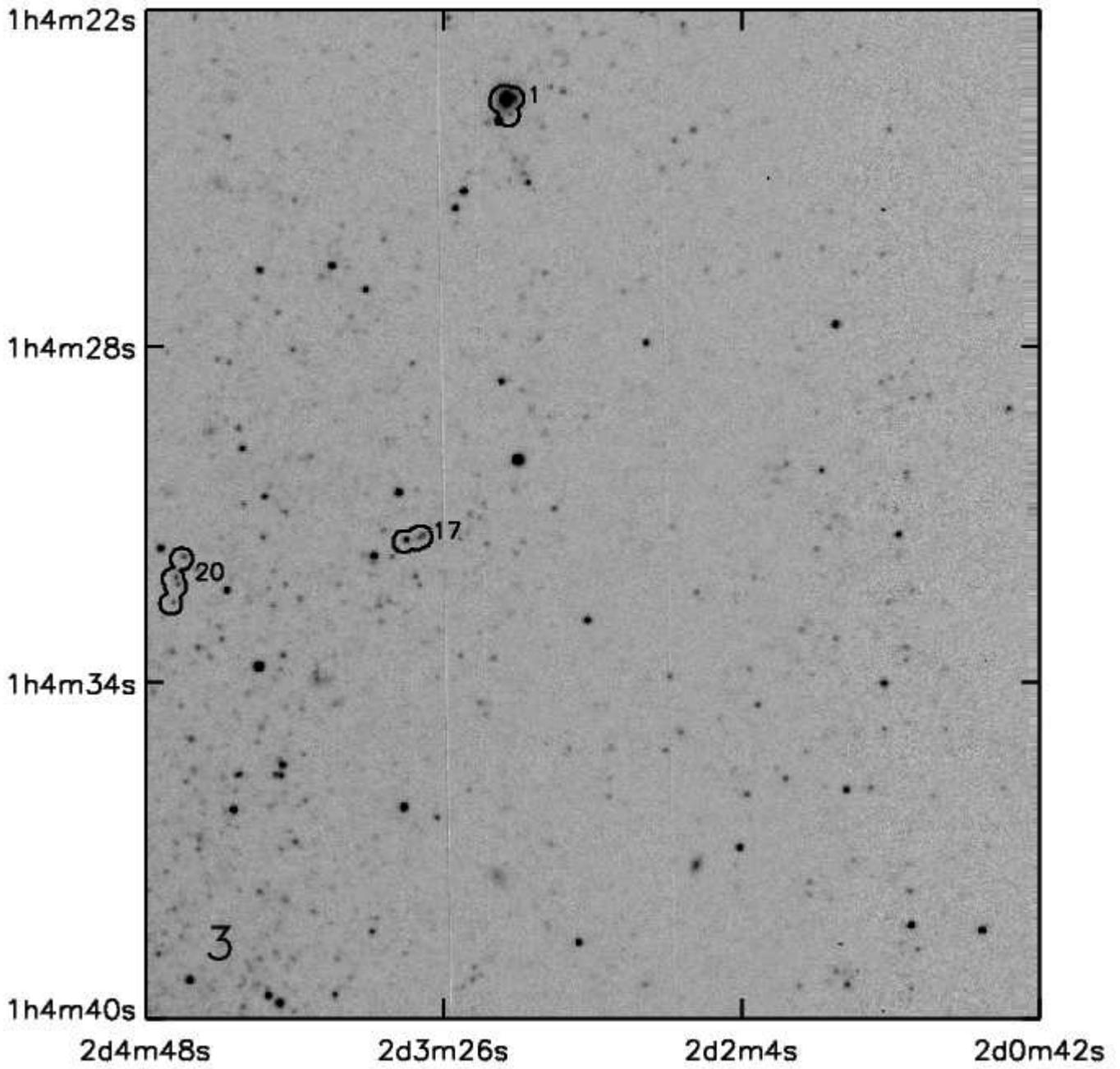}
      \caption{Same as \ref{F:f-charts1}, quadrant~3.
The enclosed associations are \#1, \#17 and \#20.
}
       \label{F:f-charts3}
   \end{figure*}

\begin{figure*}[!h]
%   \centering
%   \hspace{-1.75cm}
   \includegraphics[angle=180]{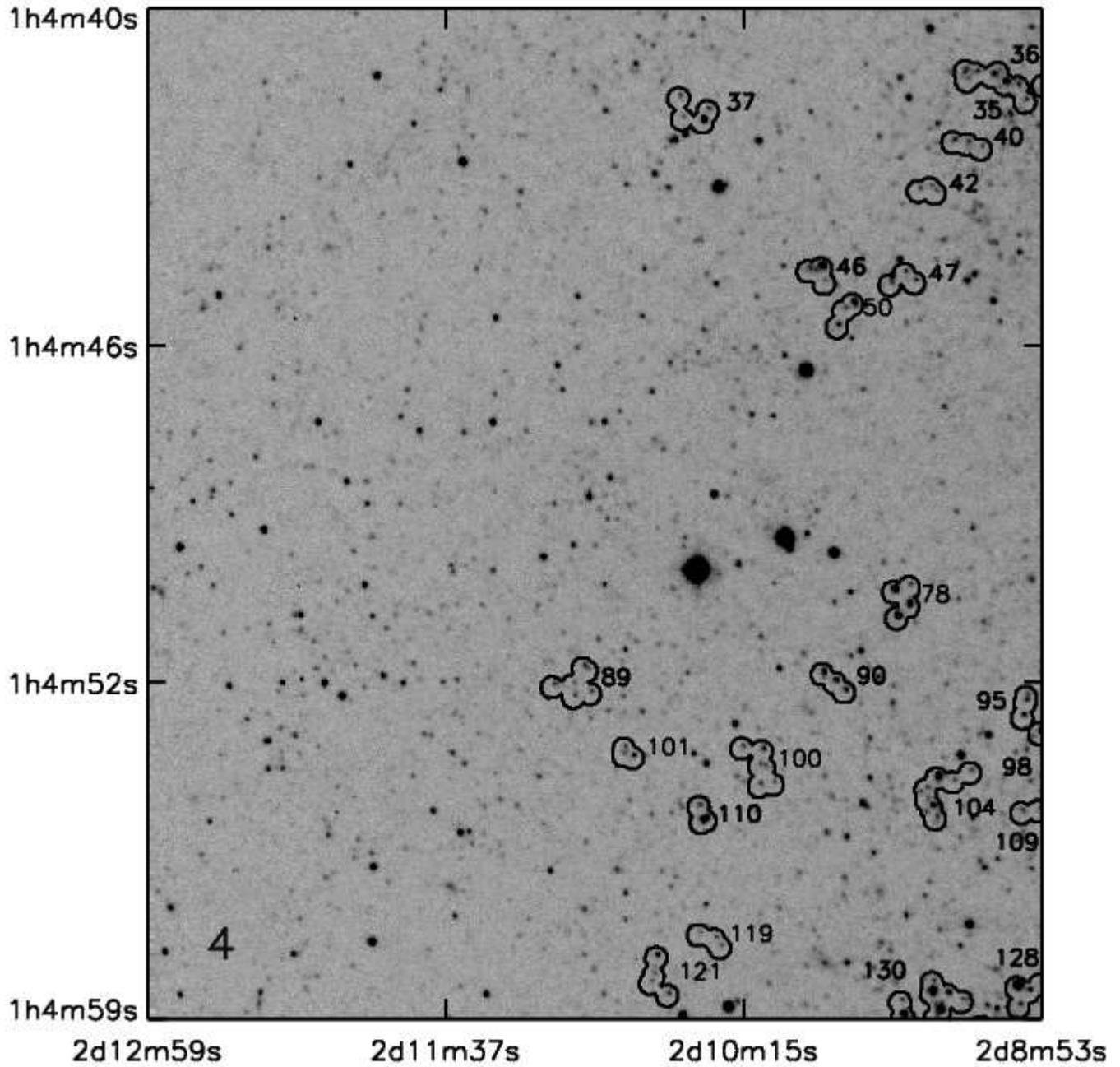}
      \caption{Same as \ref{F:f-charts1}, quadrant~4.
Associations present in this quadrant:
\#35-\#37, \#40, \#42, \#46, \#47, \#50, \#78, \#89, \#90, \#95, \#98, \#100, \#101,
\#104,\# 109\#, 110, \#119, \#121, \#127, \#128 and \#130.
}
       \label{F:f-charts4}
   \end{figure*}

\begin{figure*}[!h]
%   \centering
%   \hspace{-1.75cm}
   \includegraphics[angle=180]{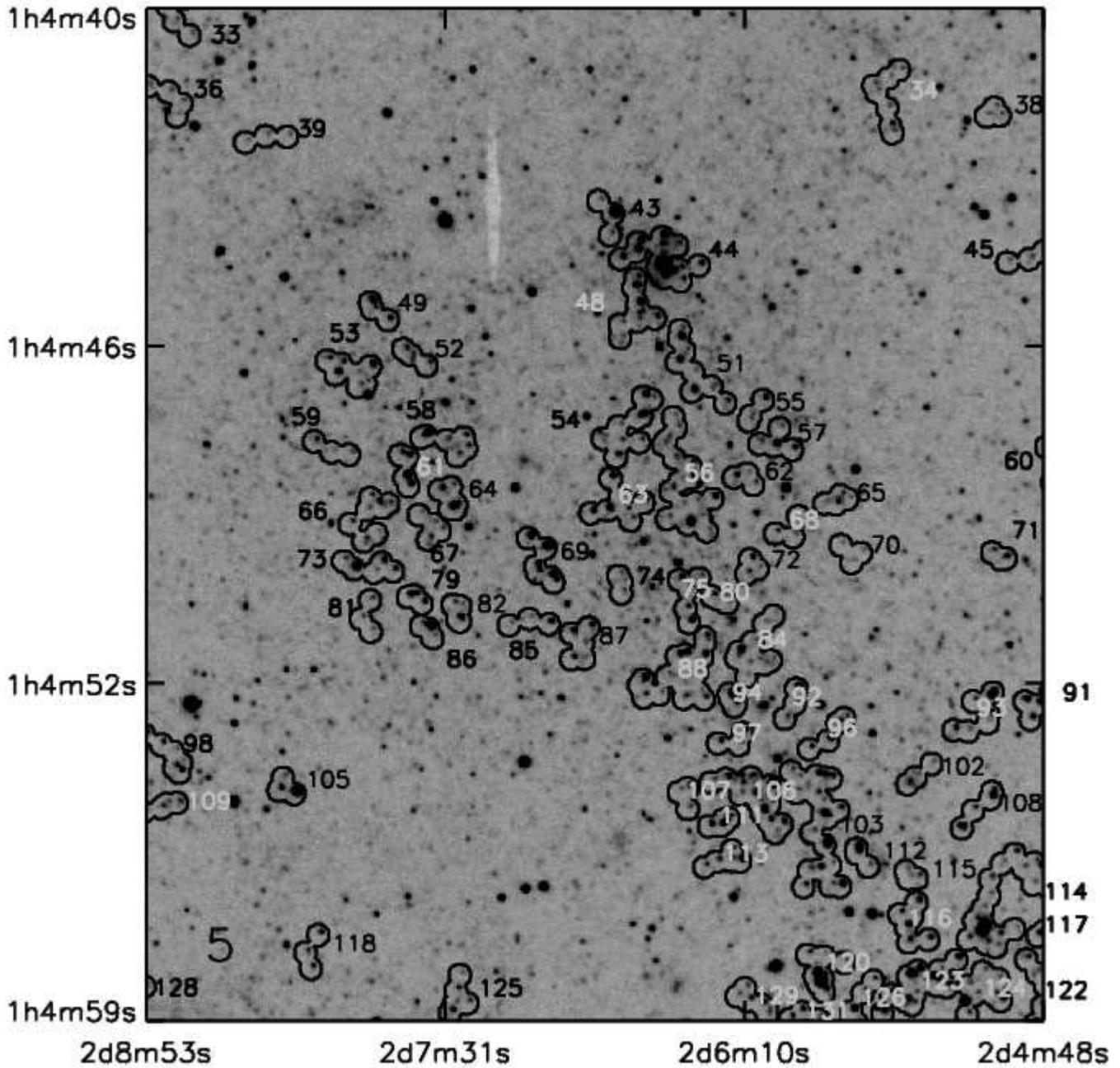}
      \caption{Same as \ref{F:f-charts1}, quadrant~5.
Associations present in this quadrant:
\#33, \#34, \#36, \#38, \#39, \#43-\#45, \#48, \#49, \#51-\#75, \#79-\#82, \#84-\#88,
\#91-\#94, \#96-\#98, \#102-\#103, \#105-\#109, \#111-\#118, \#120, \#122-\#126, \#128, \#129 and \#131.
}
       \label{F:f-charts5}
   \end{figure*}

\begin{figure*}[!h]
%   \centering
%   \hspace{-1.75cm}
   \includegraphics[angle=180]{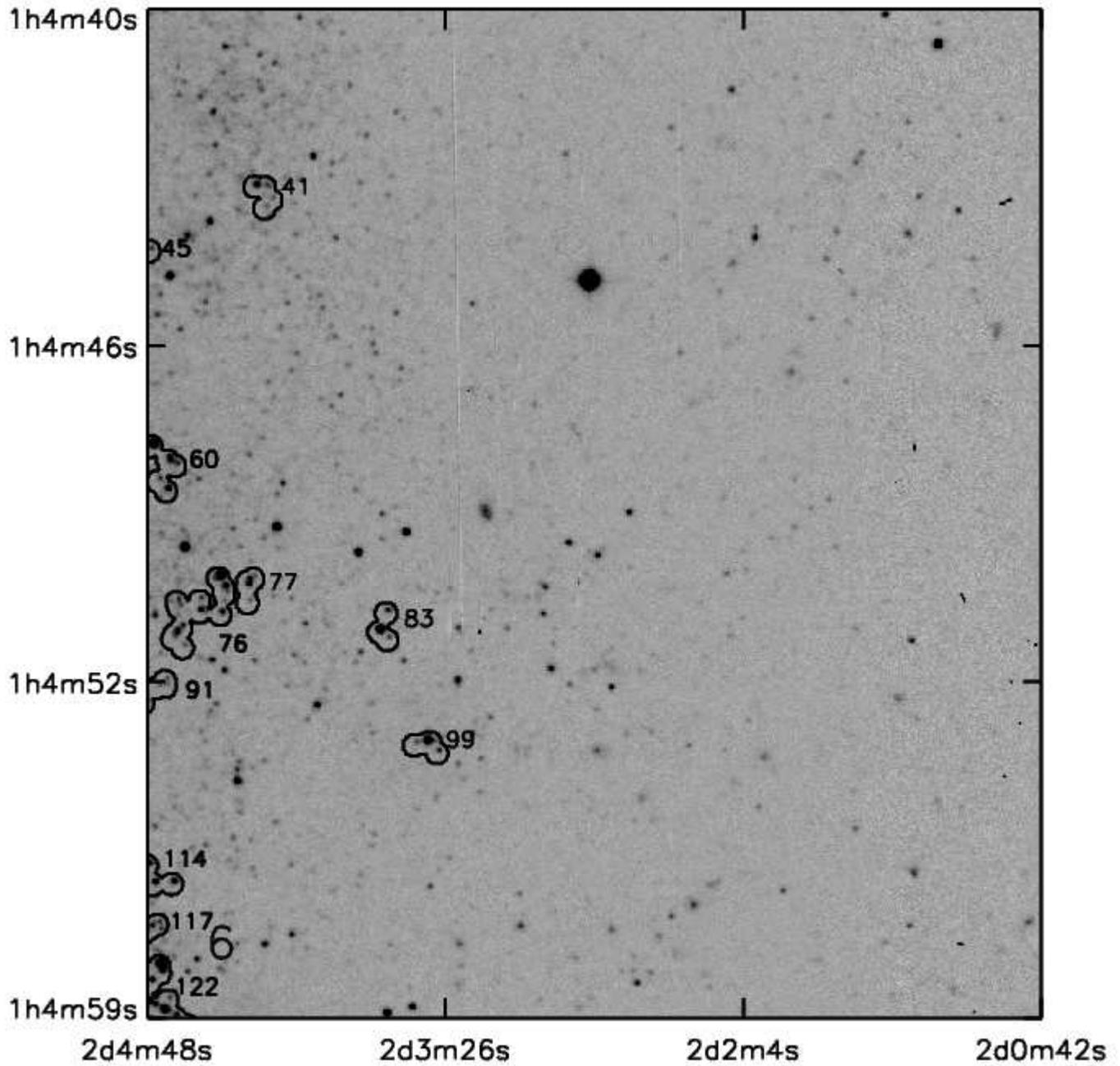}
      \caption{Same as \ref{F:f-charts1}, quadrant~6.
The enclosed associations are \#41, \#45, \#60, \#76, \#77, \#83, \#91, \#99, \#114, \#117 and \#122.}
       \label{F:f-charts6}
   \end{figure*}

\begin{figure*}[!h]
%   \centering
%   \hspace{-1.75cm}
   \includegraphics[angle=180]{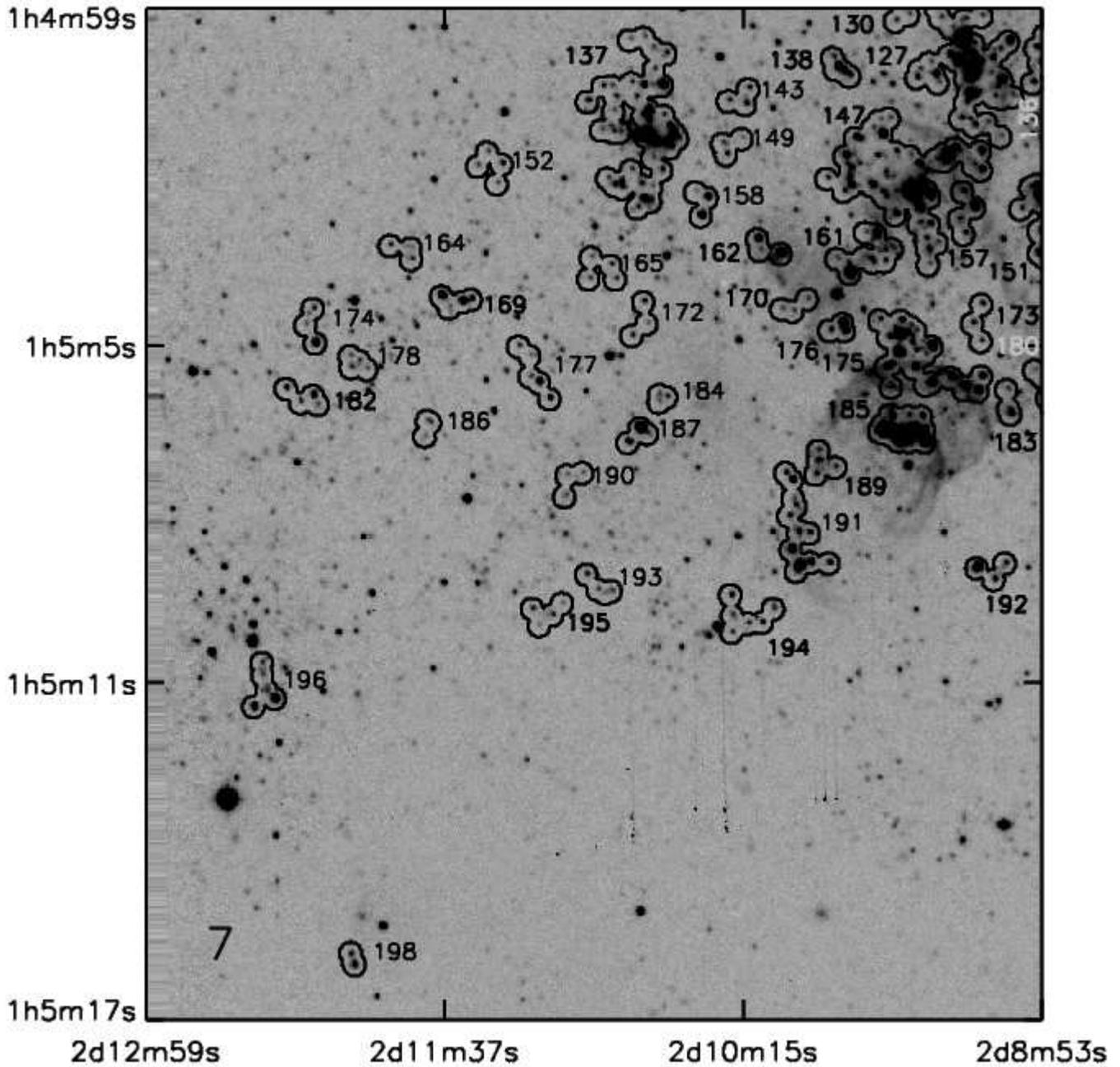}
      \caption{Same as \ref{F:f-charts1}, quadrant~7.
Associations present in this quadrant:
\#127, \#130, \#136-\#138, \#143, \#147, \#149, \#151, \#152, \#157, \#158, \#161, \#162, \#164,
\#165, \#169, \#170, \#172-\#178,
\#180, \#182-\#187, \#189-\#196 and \#198.
}
       \label{F:f-charts7}
   \end{figure*}

\begin{figure*}[!h]
%   \centering
%   \hspace{-1.75cm}
   \includegraphics[angle=180]{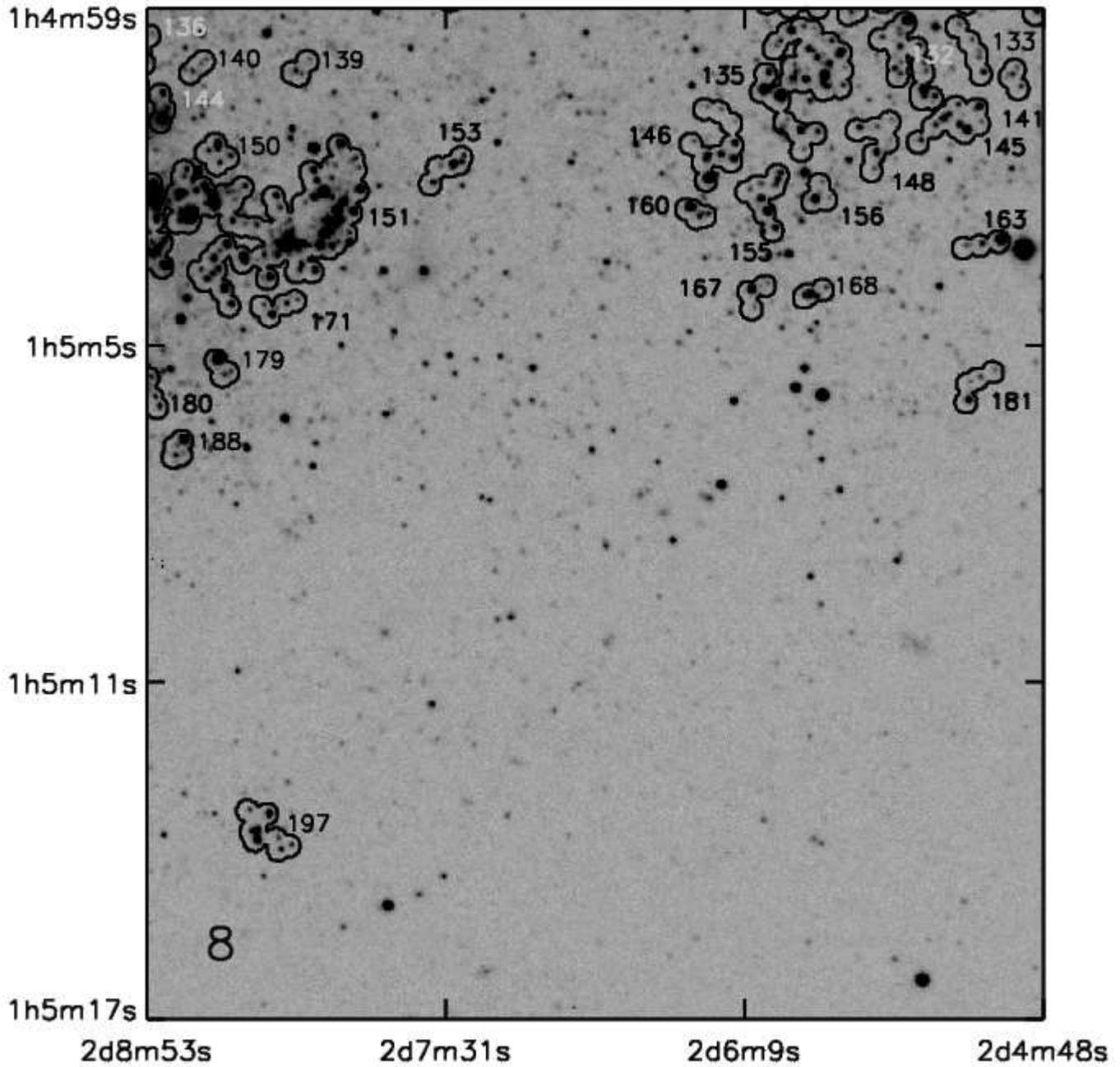}
      \caption{Same as \ref{F:f-charts1}, quadrant~8.
Associations present in this quadrant:
\#122-\#126, \#129, \#131-\#133, \#135, \#136, \#139-\#141, \#144-\#146, \#148, \#150, \#151, \#153,
\#155, \#156,\# 160, \#163, \#167, \#168, \#171, \#179-\#181, \#188 and \#197.
}
       \label{F:f-charts8}
   \end{figure*}

\begin{figure*}[!h]
%   \centering
%   \hspace{-1.75cm}
   \includegraphics[angle=180]{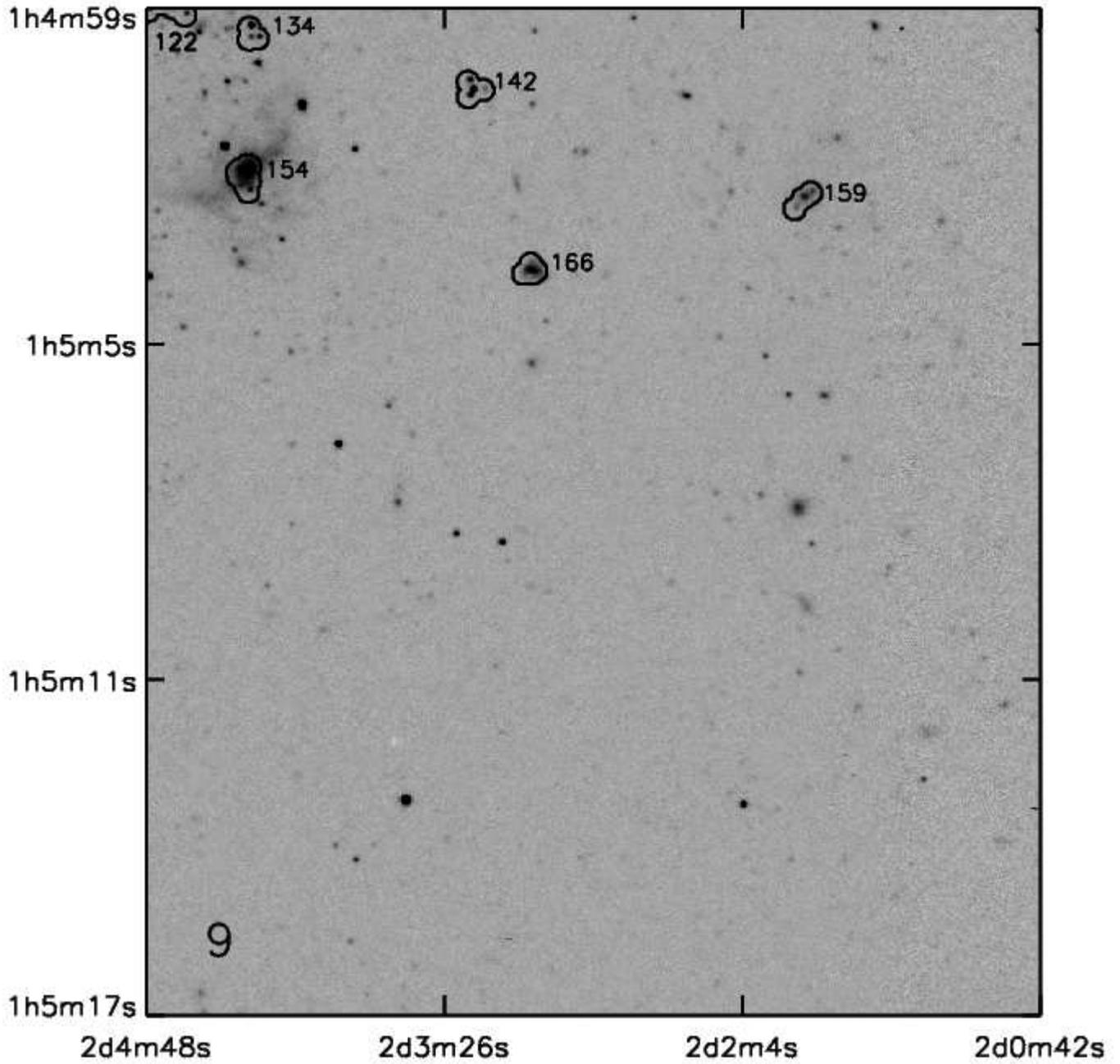}
      \caption{Same as \ref{F:f-charts1}, quadrant~9.
The enclosed associations are \#122, \#134, \#142, \#154,\# 159 and \#166.
}
       \label{F:f-charts9}
   \end{figure*}

The positions of the associations were marked on 
the galaxy in Fig.~\ref{F:all}.
Zooms into different galactic sectors are provided in this Appendix,
Figs.~\ref{F:f-charts1}-\ref{F:f-charts9}.

%%%%%%%%%%%%%%%%%%%%%%% CATALOGO CON TODAS %%%%%%%%%%%%%%%%%%%%%%%%%%%%%%%%

{\tiny

%%_____________________________________________________________
%%                                 A more than 1 page table
%%  In the preamble, use:   \usepackage{longtable}
%%-------------------------------------------------------------
%% In the text, at the place where the large table should appear
%% add the command:
%\addtocounter{table}{1}
\longtab{13}{
\begin{longtable}{ccccclc}
\caption{\label{T:ASSprop} Catalogue of OB associations in IC~1613.}\\
%\centering
%\hspace{-9cm}
\hline \hline             
ID  & RA $[\deg]$ & DEC $[\deg]$ & N$\rm_{OB}$ & N$\rm_{TOT}$ & Quad & Notes \\
    & (J2000.0)   & (J2000.0)    &             &             &      &       \\ 
(1) & (2)         & (3)          & (4)         & (5)         & (6)  & (7)   \\
\hline
\endfirsthead
\caption{continued.}\\
\hline\hline
ID  & RA $[\deg]$ & DEC $[\deg]$ & N$\rm_{OB}$ & N$\rm_{TOT}$ & Quad & Notes \\
    & (J2000.0)   & (J2000.0)    &             &             &      &       \\ 
(1) & (2)         & (3)          & (4)         & (5)         & (6)  & (7)   \\
\hline
\endhead
\hline
\endfoot
        1 &   16.096912 &    2.052357 &      8 &     10 & 3     &  Gal  \\ %
        2 &   16.102215 &    2.110539 &      3 &      3 & 2     &   -   \\ % A
        3 &   16.103825 &    2.152657 &      4 &      4 & 1     &   -   \\ % A
        4 &   16.103556 &    2.127854 &      3 &      3 & 2     &   d   \\ %
        5 &   16.108092 &    2.114839 &      3 &      3 & 2     &   ?   \\ %
        6 &   16.110625 &    2.110028 &      3 &      3 & 2     &   -   \\ % A
        7 &   16.112182 &    2.139603 &      4 &      4 & 2     &   -   \\ % Q
        8 &   16.116793 &    2.148229 &      4 &      4 & 1,2   &   -   \\ % QR
        9 &   16.116749 &    2.181977 &      5 &      5 & 1     &   -   \\ % 
       10 &   16.120176 &    2.176666 &      3 &      3 & 1     &   n   \\ % R
       11 &   16.125909 &    2.149869 &      4 &      4 & 1     &   -   \\ % A
       12 &   16.125697 &    2.104176 &      3 &      3 & 2     &   d   \\ % 
       13 &   16.127069 &    2.128816 &      4 &      4 & 2     &   -   \\ % SA
       14 &   16.129429 &    2.097499 &      4 &      4 & 2     &   -   \\ % 
       15 &   16.130444 &    2.093961 &      3 &      3 & 2     &   -   \\ % 
       16 &   16.133370 &    2.165873 &     10 &     12 & 1     &   n   \\ % 
       17 &   16.130397 &    2.059373 &      3 &      3 & 3     &   -   \\ % 
       18 &   16.132333 &    2.150075 &      4 &      4 & 1,2   &   -   \\ % 
       19 &   16.132298 &    2.169800 &      3 &      4 & 1     &   ?   \\ % 
       20 &   16.133645 &    2.077595 &      4 &      4 & 3     &   -   \\ % 
       21 &   16.135528 &    2.161271 &     17 &     18 & 1     &   n   \\ % 
       22 &   16.138624 &    2.161875 &      3 &      3 & 1     &   -   \\ % 
       23 &   16.140158 &    2.098123 &      4 &      4 & 2     &   ?   \\ % Q
       24 &   16.143162 &    2.105718 &      5 &      5 & 2     &   -   \\ % 
       25 &   16.150666 &    2.098182 &      7 &      9 & 2     &   d   \\ % Q
       26 &   16.149934 &    2.104582 &      3 &      3 & 2     &   -   \\ % S
       27 &   16.150782 &    2.094274 &      5 &      5 & 2     &   n   \\ % 
       28 &   16.154294 &    2.093873 &      4 &      4 & 2     &   n   \\ %
       29 &   16.156629 &    2.107164 &      3 &      3 & 2     &   ?   \\ % Q
       30 &   16.157865 &    2.118325 &      4 &      4 & 2     &   -   \\ % Q
       31 &   16.160063 &    2.086166 &      3 &      4 & 2     &   -   \\ % A
       32 &   16.165508 &    2.129537 &      7 &      7 & 2     &  Gal  \\ % 
       33 &   16.167062 &    2.145645 &      5 &      7 & 2,5   &   -   \\ % Q
       34 &   16.173860 &    2.091680 &      7 &      8 & 5     &   -   \\ % B
       35 &   16.172452 &    2.151775 &      8 &      9 & 4     &   -   \\ % SR
       36 &   16.173835 &    2.146373 &      4 &      7 & 4,5   &   -   \\ % 
       37 &   16.174840 &    2.174687 &      4 &      5 & 4     &   -   \\ % Q
       38 &   16.174955 &    2.083569 &      3 &      3 & 5     &   -   \\ % 
       39 &   16.176810 &    2.138951 &      3 &      3 & 5     &   d   \\ % 
       40 &   16.177356 &    2.153791 &      3 &      3 & 4     &   ?   \\ % QB
       41 &   16.181246 &    2.070946 &      4 &      4 & 6     &   -   \\ % A
       42 &   16.180837 &    2.156761 &      3 &      3 & 4     &   d   \\ % 
       43 &   16.182817 &    2.112977 &      3 &      5 & 5     &   n   \\ % SA
       44 &   16.186156 &    2.108783 &     16 &     17 & 5     &   -   \\ % QBS
       45 &   16.185969 &    2.081011 &      3 &      3 & 5,6   &   ?   \\ % Q
       46 &   16.187206 &    2.165141 &      3 &      3 & 4     &   d   \\ % 
       47 &   16.187683 &    2.158649 &      3 &      5 & 4     &   ?   \\ % RBV
       48 &   16.190017 &    2.110850 &      9 &      9 & 5     &   n   \\ % R
       49 &   16.189936 &    2.130523 &      3 &      5 & 5     &   d   \\ % 
       50 &   16.190243 &    2.163109 &      3 &      5 & 4     &   d   \\ % 
       51 &   16.194281 &    2.106449 &      8 &     10 & 5     &   n   \\ % QR
       52 &   16.193448 &    2.127703 &      3 &      4 & 5     &   d   \\ % 
       53 &   16.194693 &    2.132499 &      7 &      8 & 5     &   n   \\ % QR
       54 &   16.198802 &    2.111230 &      9 &     10 & 5     &   n   \\ % QB
       55 &   16.197736 &    2.101776 &      3 &      3 & 5     &   d   \\ % 
       56 &   16.203010 &    2.107146 &     16 &     19 & 5     &   -   \\ % QRBA
       57 &   16.200013 &    2.100085 &      4 &      4 & 5     &   ?   \\ % R
       58 &   16.200024 &    2.125184 &      6 &      6 & 5     &   ?   \\ % VA
       59 &   16.200528 &    2.133959 &      3 &      4 & 5     &   d   \\ % 
       60 &   16.202105 &    2.078538 &      6 &      6 & 5,6   &   ?   \\ % SQB
       61 &   16.202044 &    2.128357 &      4 &      4 & 5     &   n   \\ % BA
       62 &   16.202794 &    2.102484 &      3 &      3 & 5     &   d   \\ %
       63 &   16.204786 &    2.112106 &      7 &      7 & 5     &   -   \\ % F
       64 &   16.204397 &    2.124905 &      5 &      5 & 5     &   -   \\ % QA
       65 &   16.204474 &    2.095443 &      4 &      5 & 5     &   d   \\ % 
       66 &   16.205571 &    2.130890 &      8 &     10 & 5     &   ?   \\ % QR
       67 &   16.206485 &    2.126462 &      4 &      5 & 5     &   ?   \\ % R
       68 &   16.206628 &    2.099115 &      3 &      3 & 5     &   ?   \\ % FA
       69 &   16.209295 &    2.117948 &      6 &      8 & 5     &   n   \\ % 
       70 &   16.208526 &    2.094393 &      3 &      3 & 5     &   d   \\ % 
       71 &   16.208783 &    2.083076 &      3 &      4 & 5     &   d   \\ % 
       72 &   16.209589 &    2.101978 &      3 &      3 & 5     &   ?   \\ % FV
       73 &   16.209537 &    2.131091 &      5 &      5 & 5     &   -   \\ % R
       74 &   16.210802 &    2.111945 &      3 &      3 & 5     &   d   \\ % 
       75 &   16.211725 &    2.106787 &      5 &      8 & 5     &   ?n  \\ % 
       76 &   16.213125 &    2.076123 &     12 &     13 & 6     &   -   \\ % 
       77 &   16.211226 &    2.072083 &      4 &      4 & 6     &   ?   \\ % QSAV
       78 &   16.212151 &    2.158723 &      4 &      4 & 4     &   ?   \\ % QV
       79 &   16.211941 &    2.127620 &      3 &      3 & 5     &   ?   \\ % S
       80 &   16.212032 &    2.104437 &      3 &      5 & 5     &   -   \\ % SA
       81 &   16.213312 &    2.131232 &      3 &      3 & 5     &   -   \\ % V
       82 &   16.213018 &    2.124339 &      4 &      5 & 5     &   ?   \\ % V
       83 &   16.214367 &    2.061752 &      3 &      3 & 6     &   n   \\ % Q
       84 &   16.216014 &    2.101695 &     10 &     10 & 5     &   -   \\ % ASBF
       85 &   16.213772 &    2.118912 &      3 &      3 & 5     &   ?n  \\ % BF
       86 &   16.214380 &    2.126652 &      3 &      3 & 5     &   ?n  \\ % BF
       87 &   16.215442 &    2.115048 &      5 &      5 & 5     &   -   \\ % BA
       88 &   16.217727 &    2.107367 &     16 &     17 & 5     &   -   \\ % BF
       89 &   16.218450 &    2.183555 &      6 &      6 & 4     &   -   \\ % V
       90 &   16.218416 &    2.163930 &      3 &      3 & 4     &   ?   \\ % 
       91 &   16.219528 &    2.079689 &      6 &      6 & 5,6   &   -   \\ % ST
       92 &   16.219961 &    2.098888 &      3 &      3 & 5     &   d   \\ % 
       93 &   16.220890 &    2.084779 &      6 &      7 & 5     &   -   \\ % BS
       94 &   16.219702 &    2.103431 &      4 &      4 & 5     &   ?   \\ % B
       95 &   16.220188 &    2.149358 &      3 &      3 & 4     &   -   \\ % 
       96 &   16.222186 &    2.096009 &      4 &      5 & 5     &   n   \\ % 
       97 &   16.222757 &    2.103359 &      3 &      3 & 5     &   n   \\ % S
       98 &   16.223875 &    2.146328 &      6 &      7 & 4,5   &   n   \\ % RV
       99 &   16.223379 &    2.058472 &      4 &      4 & 6     &   n   \\ % V
      100 &   16.224860 &    2.169550 &      6 &      6 & 4     &   -   \\ % QSB
      101 &   16.223822 &    2.179711 &      3 &      3 & 4     &   d   \\ % 
      102 &   16.225417 &    2.089271 &      3 &      3 & 5     &   d   \\ % 
      103 &   16.229149 &    2.096880 &     19 &     23 & 5     &   n   \\ % AFRQSB
      104 &   16.226815 &    2.155823 &      7 &     10 & 4     &   ?n  \\ % VF
      105 &   16.226303 &    2.137465 &      3 &      3 & 5     &   dn  \\ % 
      106 &   16.227180 &    2.101775 &     14 &     18 & 5     &   n   \\ % FBS
      107 &   16.227032 &    2.107051 &      3 &      3 & 5     &   d   \\ % 
      108 &   16.227880 &    2.084607 &      4 &      6 & 5     &   ?   \\ % B
      109 &   16.227894 &    2.147403 &      5 &      5 & 4,5   &   ?   \\ % QF
      110 &   16.228626 &    2.173969 &      3 &      4 & 4     &   ?   \\ % 
      111 &   16.229030 &    2.104253 &      5 &      5 & 5     &   ?   \\ % BRSA
      112 &   16.231603 &    2.093674 &      3 &      4 & 5     &   ?   \\ % BRV
      113 &   16.231901 &    2.103999 &      5 &      6 & 5     &   -   \\ % B
      114 &   16.235155 &    2.082594 &     22 &     23 & 5,6   &   n   \\ % QBA
      115 &   16.233123 &    2.089860 &      3 &      3 & 5     &   d   \\ % 
      116 &   16.236758 &    2.089753 &      6 &      7 & 5     &   n   \\ % Q
      117 &   16.237337 &    2.079615 &      3 &      4 & 5,6   &   n   \\ % B
      118 &   16.238855 &    2.135455 &      3 &      3 & 5     &   d   \\ % 
      119 &   16.238087 &    2.173274 &      3 &      3 & 4     &   -   \\ % 
      120 &   16.240415 &    2.096584 &      7 &      7 & 5     &   n   \\ % B
      121 &   16.240945 &    2.177397 &      4 &      4 & 4     &   -   \\ % 
      122 &   16.242760 &    2.079217 &     15 &     19 &5,6,8,9&   n   \\ % QR
      123 &   16.241480 &    2.089006 &     14 &     15 & 5,8   &   -   \\ % SQ
      124 &   16.241957 &    2.084267 &      7 &      8 & 5,8   &   n   \\ % VQ
      125 &   16.242312 &    2.124345 &      5 &      5 & 5,8   &   n   \\ % QR
      126 &   16.243098 &    2.093806 &      5 &      5 & 5,8   &   n   \\ % QV
      127 &   16.248152 &    2.153727 &     49 &     54 & 4,7   &   n   \\ % AVBQ
      128 &   16.242096 &    2.149247 &      4 &      5 & 4,5   &   n   \\ % AVQ
      129 &   16.242767 &    2.101736 &      6 &      6 & 5,8   &   -   \\ % SRQ
      130 &   16.244092 &    2.158721 &      4 &      5 & 4,7   &   -   \\ % RV
      131 &   16.243844 &    2.098030 &      3 &      3 & 5,8   &   -   \\ % 
      132 &   16.248173 &    2.090002 &     13 &     14 & 8     &   -   \\ % RQBSA
      133 &   16.247413 &    2.085148 &      5 &      5 & 8     &   -   \\ % A
      134 &   16.246209 &    2.071744 &      3 &      3 & 9     &   n   \\ % Q
      135 &   16.249324 &    2.097876 &     31 &     37 & 8     &   -   \\ % ASRBQ
      136 &   16.247844 &    2.148337 &      4 &      4 & 7,8   &   n   \\ % QFA
      137 &   16.253227 &    2.178541 &     46 &     48 & 7     &   n   \\ % BSQ
      138 &   16.248630 &    2.163478 &      6 &      6 & 7     &   ?   \\ % QVF
      139 &   16.248756 &    2.136402 &      3 &      3 & 8     &   d   \\ % 
      140 &   16.248518 &    2.144179 &      3 &      3 & 8     &   ?n  \\ % BA
      141 &   16.249494 &    2.081922 &      3 &      3 & 8     &  Gal  \\ % 
      142 &   16.250423 &    2.054868 &      4 &      4 & 9     &   -   \\ % 
      143 &   16.250916 &    2.170979 &      3 &      3 & 7     &   -   \\ % A
      144 &   16.251720 &    2.146901 &      3 &      3 & 8     &   n   \\ % Q
      145 &   16.252874 &    2.086635 &      9 &      9 & 8     &   -   \\ % SQBA
      146 &   16.254549 &    2.104795 &     10 &     10 & 8     &   n   \\ % ASQ
      147 &   16.257084 &    2.158485 &     53 &     58 & 7     &   n   \\ % ABRQ
      148 &   16.254275 &    2.092485 &      6 &      7 & 8     &   n   \\ % Q
      149 &   16.254539 &    2.171786 &      3 &      3 & 7     &   ?n  \\ % FB
      150 &   16.255150 &    2.142742 &      5 &      6 & 8     &   -   \\ % Q
      151 &   16.260516 &    2.140498 &     99 &    103 & 7,8   &   n   \\ % ASFQ
      152 &   16.256125 &    2.190009 &      4 &      4 & 7     &   -   \\ % 
      153 &   16.256249 &    2.125097 &      5 &      5 & 8     &   -   \\ % AB
      154 &   16.256969 &    2.072318 &     11 &     12 & 9     &   n   \\ % AQ
      155 &   16.258348 &    2.100803 &      9 &     10 & 8     &   -   \\ % QF
      156 &   16.258377 &    2.096784 &      3 &      5 & 8     &   d   \\ % 
      157 &   16.259700 &    2.153904 &      4 &      4 & 7     &   ?n  \\ % B
      158 &   16.258826 &    2.173986 &      3 &      3 & 7     &   ?n  \\ % BF, LBV
      159 &   16.258869 &    2.029636 &      3 &      3 & 9     &  Gal  \\ % 
      160 &   16.259788 &    2.106187 &      5 &      5 & 8     &   -   \\ % ARBF
      161 &   16.262861 &    2.161338 &     11 &     12 & 7     &   n   \\ % ABSQF
      162 &   16.262323 &    2.169053 &      3 &      5 & 7     &   n   \\ % V
      163 &   16.262203 &    2.084492 &      3 &      3 & 8     &   -   \\ % S
      164 &   16.262549 &    2.196617 &      3 &      4 & 7     &   ?   \\ % B
      165 &   16.264022 &    2.181578 &      4 &      4 & 7     &   -   \\ % 
      166 &   16.264283 &    2.050493 &      6 &      6 & 9     &  Gal  \\ % 
      167 &   16.266025 &    2.101659 &      3 &      3 & 8     &   ?   \\ % SFB
      168 &   16.265918 &    2.097347 &      3 &      4 & 8     &   ?   \\ % B
      169 &   16.266384 &    2.192655 &      4 &      4 & 7     &   -   \\ % 
      170 &   16.266803 &    2.166934 &      3 &      3 & 7     &   ?   \\ % QBA
      171 &   16.266873 &    2.138073 &      4 &      5 & 8     &   -   \\ % 
      172 &   16.267850 &    2.178595 &      3 &      3 & 7     &   d   \\ % 
      173 &   16.268078 &    2.152881 &      3 &      3 & 7     &   -   \\ % BR
      174 &   16.268225 &    2.203677 &      3 &      3 & 7     &   -   \\ % 
      175 &   16.270523 &    2.157452 &     28 &     30 & 7     &   n   \\ % RBSQ
      176 &   16.268470 &    2.163499 &      3 &      3 & 7     &   n   \\ % B
      177 &   16.271771 &    2.186759 &      5 &      5 & 7     &   -   \\ % R
      178 &   16.271105 &    2.200210 &      4 &      6 & 7     &   -   \\ % SA
      179 &   16.271636 &    2.142294 &      3 &      3 & 8     &   ?n  \\ % BT
      180 &   16.273082 &    2.147837 &      4 &      7 & 7,8   &   -   \\ % BS
      181 &   16.272820 &    2.084834 &      4 &      4 & 8     &   d   \\ % 
      182 &   16.273732 &    2.204206 &      4 &      4 & 7     &   -   \\ % 
      183 &   16.274270 &    2.150566 &      3 &      4 & 7     &   d   \\ % 
      184 &   16.273774 &    2.177004 &      3 &      3 & 7     &  ?n   \\ % A
      185 &   16.276164 &    2.158762 &     17 &     17 & 7     &   n   \\ % SQB
      186 &   16.275929 &    2.194830 &      3 &      3 & 7     &   ?   \\ % 
      187 &   16.276569 &    2.178761 &      3 &      3 & 7     &   n   \\ % SB
      188 &   16.277759 &    2.145682 &      4 &      5 & 8     &   ?   \\ % 
      189 &   16.278765 &    2.164795 &      4 &      4 & 7     &   ?n  \\ % ASQ
      190 &   16.280133 &    2.183874 &      3 &      3 & 7     &   d   \\ % 
      191 &   16.283802 &    2.166530 &     13 &     13 & 7     &   n   \\ % QB
      192 &   16.286930 &    2.151835 &      3 &      3 & 7     &   ?   \\ % QRV
      193 &   16.288172 &    2.181669 &      4 &      4 & 7     &   ?   \\ % 
      194 &   16.290408 &    2.170542 &      6 &      7 & 7     &   ?   \\ % AB
      195 &   16.290153 &    2.185749 &      4 &      4 & 7     &   ?   \\ % SB
      196 &   16.295735 &    2.207267 &      5 &      5 & 7     &   n   \\ % SQB
      197 &   16.306976 &    2.139009 &      8 &      8 & 8     &   -   \\ % ASQB
      198 &   16.316533 &    2.200474 &      3 &      3 & 7     &   d   \\ % 
%
%
% A= notes on age for Mup determination
% B= one or more blends
% R= sneaked in star or does not follow reddening law
% Q= members with too blue Q
% S= scatter in ages, age spread, dispersion in CMD or QV diagrams
% F= faint stars are fuzzy
% V= resolved in VIMOS images or WFPC2, but not in INT.
% T= stars have tails in INT that are resolved in VIMOS into blue+red blends

%\hline
\end{longtable}
}
%
          % label {T:OBass}
}

%%%%%%%%%%%%%%%%%%%%%%% TABLA CON MIEMBROS DE CADA ASSOC %%%%%%%%%%%%%%%%%%%%%%%%%%%%%%%%

{\tiny

%_____________________________________________________________
%                                 A rotated Table in landscape  
%  In the preamble, use:   \usepackage{lscape}
%-------------------------------------------------------------
\begin{landscape}
\begin{table*}
\caption{Members of association \#16.}\label{T:OBass}
%\centering
\hspace{-6cm}
\begin{tabular}{cccccccccccccccc}
\hline\hline             
  \multicolumn{16}{l}{\it OB-stars in Association}\\ 
\hline
ID        &RA $\rm [deg]$      &DEC $\rm [deg]$      &V      & $\sigma$-V   &B      &$\sigma$-B       &I      &$\sigma$-I       &R      &$\sigma$-R      &U       &$\sigma$-U      &$\rm \chi $      & \#      & ID\_2MASS        \\
          & (J2000.0)          & (J2000.0)           &       &              &       &                 &       &                 &       &                &        &                &                 &         &                  \\
(1)   & (2)       & (3)       & (4)     & (5)     & (6)     & (7)     & (8)     & (9)     & (10)    & (11)    & (12)    & (13)    & (14)    & (15)    &(16)            \\
\hline
 17183&  16.129984&   2.165500&   19.885&   0.0074&   19.827&   0.0055&   19.894&   0.0143&   19.936&   0.0089&   19.220&   0.0121&    1.405&    0.051& --               \\
 17397&  16.130867&   2.166583&   22.656&   0.0477&   22.636&   0.0345&   23.124&   0.1523&   22.808&   0.0642&   21.689&   0.0398&    1.219&    0.531& --               \\
 17512&  16.131291&   2.166572&   21.861&   0.0257&   21.716&   0.0182&   22.190&   0.0629&   21.964&   0.0351&   20.852&   0.0195&    1.244&    0.322& --               \\
 17896&  16.132740&   2.166766&   21.819&   0.0222&   21.655&   0.0162&   21.958&   0.0546&   21.987&   0.0335&   20.550&   0.0181&    1.200&    0.271& --               \\
 17945&  16.132886&   2.166397&   22.193&   0.0325&   22.129&   0.0245&   21.988&   0.0672&   22.139&   0.0372&   21.379&   0.0335&    1.259&    0.429& --               \\
 18051&  16.133337&   2.165283&   22.313&   0.0301&   22.172&   0.0209&   22.379&   0.0962&   22.485&   0.0511&   21.324&   0.0289&    1.193&    0.380& --               \\
 18159&  16.133736&   2.164780&   21.968&   0.0357&   21.943&   0.0200&   21.942&   0.0957&   22.053&   0.0567&   21.095&   0.0286&    1.629&    0.776& --               \\
 18444&  16.134765&   2.165754&   22.547&   0.0351&   22.455&   0.0254&   22.586&   0.1204&   22.652&   0.0590&   21.678&   0.0350&    1.158&    0.472& --               \\
 18867&  16.136389&   2.165391&   20.818&   0.0132&   20.641&   0.0090&   21.012&   0.0270&   20.981&   0.0164&   19.598&   0.0139&    1.427&    0.226& --               \\
 19245&  16.137705&   2.165704&   22.499&   0.0302&   22.462&   0.0236&   22.531&   0.0812&   22.493&   0.0416&   21.912&   0.0416&    0.999&    0.037& --               \\

\hline
%\hline
  \multicolumn{16}{l}{~~}\\ 
  \multicolumn{16}{l}{\it Non OB-stars in Association}\\ 
\hline
ID        &RA $\rm [deg]$      &DEC $\rm [deg]$      &V      & $\sigma$-V   &B      &$\sigma$-B       &I      &$\sigma$-I       &R      &$\sigma$-R      &U       &$\sigma$-U      &$\rm \chi $      & \#           & ID\_2MASS        \\
          & (J2000.0)          & (J2000.0)           &       &              &       &                 &       &                 &       &                &        &                &                 &         &                  \\
(1)   & (2)       & (3)       & (4)     & (5)     & (6)     & (7)     & (8)     & (9)     & (10)    & (11)    & (12)    & (13)    & (14)    & (15)    &(16)           \\
\hline
 16825&  16.128596&   2.164702&   21.332&   0.0173&   21.617&   0.0150&   20.855&   0.0280&   21.120&   0.0200&   21.678&   0.0367&    1.266&    0.383& --               \\
 18116&  16.133573&   2.166236&   21.772&   0.0352&   22.151&   0.0217&   21.064&   0.0657&   21.464&   0.0485&   22.176&   0.0466&    1.789&    0.661& --               \\

\hline
\end{tabular}
\end{table*}
\end{landscape}
%
   % label {T:ASSprop}
}

%%%%%%%%%%%%%%%%%%% FINDING CHARTS %%%%%%%%%%%%%%%%%%%%%%%%%%%%%%%%%%%%%%%%%%%%%%%%%%

\end{appendix}

\end{document}